\documentclass{article} 
\usepackage{grad_preprint,times}

\usepackage{amsmath,amsfonts,bm}

\def\eqref#1{equation~\ref{#1}}

\def\1{\bm{1}}

\DeclareMathAlphabet{\mathsfit}{\encodingdefault}{\sfdefault}{m}{sl}
\SetMathAlphabet{\mathsfit}{bold}{\encodingdefault}{\sfdefault}{bx}{n}

\usepackage{hyperref}
\usepackage{url}

\usepackage{tabularx}
\usepackage{multirow}
\usepackage{placeins}
\usepackage[ruled,vlined]{algorithm2e} 
\usepackage{amsmath} 

\usepackage{booktabs}       
\usepackage{siunitx}        
\usepackage{threeparttable} 
\usepackage{caption}        
\usepackage{wrapfig,epsfig}
\usepackage{mathtools}
\usepackage{dsfont}
\usepackage{graphicx}
\usepackage{nomencl}
\usepackage{multirow}
\usepackage{makecell}
\usepackage{adjustbox}
\usepackage{subcaption}
\usepackage{enumerate}
\usepackage{enumitem}
\usepackage{multirow}
\usepackage{amssymb}
\usepackage[symbol]{footmisc}
\usepackage{colortbl}
\usepackage{diagbox}
\usepackage{fontawesome}

\usepackage{booktabs}
\usepackage{bbding}
\usepackage{pifont}
\usepackage{wasysym}
\usepackage{utfsym}

\usepackage{xcolor}
\definecolor{bestcolor}{HTML}{9898FF}
\definecolor{mygreen}{RGB}{0,150,70}
\definecolor{myred}{RGB}{201,35,35}
\usepackage{needspace}
\makeatletter
\newcommand{\WrapTableClear}{\par
  \ifnum\c@WF@wrappedlines>1
    \vspace{\dimexpr\c@WF@wrappedlines\baselineskip-\baselineskip\relax}
  \fi
  \WFclear}
\makeatother

\newcommand{\ResultCell}[1]{\ensuremath{\csname result@#1\endcsname}}

\newcommand{\FinalStreamProtocol}{five seeds and two task orders on the Decomp/MCP/MRT and family-isolated mechanism streams; implementation details in Appendix~L}
\newcommand{\FinalRoutingResults}{The three-snapshot pool occupies 7.74\,MB.}

\expandafter\def\csname result@det512-decomp-tpr\endcsname{95.4}
\expandafter\def\csname result@det512-decomp-fpr\endcsname{13.8}
\expandafter\def\csname result@det512-mcp-tpr\endcsname{52.0}
\expandafter\def\csname result@det512-mcp-fpr\endcsname{14.1}
\expandafter\def\csname result@det512-mrt-tpr\endcsname{83.3}
\expandafter\def\csname result@det512-mrt-fpr\endcsname{58.0}

\title{GRAD: Graph-Structured Antibody Defense for LLM Agents}

\author{%
Tianqi Song$^{1}$ \quad Jiawei Du$^{2}$ \quad Xuemei Jia$^{1}$ \quad Zheng Wang$^{1}$\\
$^{1}$Wuhan University\\
$^{2}$Agency for Science, Technology and Research (A*STAR)\\
\texttt{2024302111167@whu.edu.cn}}

\iclrfinalcopy
\hypersetup{hidelinks,pdfauthor={Tianqi Song, Jiawei Du, Xuemei Jia, Zheng Wang},pdftitle={GRAD: Graph-Structured Antibody Defense for LLM Agents}}
\begin{document}

\maketitle

\begin{abstract}
Memory poisoning and decomposition attacks distribute malicious evidence
across the operations of large language model agents.
Existing request-level defenses can miss such attacks because they assess
individual inputs without connecting earlier information to later actions.
Moreover, detectors must adapt as task contexts change, while updates
to shared parameters can impair previously learned detection capabilities.
To address these challenges, we propose \textbf{GRAD}, which combines
graph-based evidence aggregation across operations with an extensible
collection of detection experts.
Its \textbf{Semantic--Structural Understanding Module (SSUM)} jointly encodes
operation text, event attributes, and typed candidate relations in an execution graph
to aggregate distributed attack evidence into a graph-level risk score.
Its \textbf{Continual Antibody Adaptation (CAA)} adds experts on frozen
encoder snapshots and routes inputs by observable graph signatures,
acquiring and reusing detection capabilities as tasks arrive.
Across Decomp, MCP, and MRT, GRAD achieves AUROCs of 97.19\%,
77.59\%, and 86.60\%, respectively, outperforming Sequential Monitor and locally adapted Agent-Sentry
and NeuroTaint. CAA improves final accuracy by
13.08 percentage points over Replay on the
cross-dataset task stream. The gain remains 9.05 points with equal
family/base-case weights within each task. Its compact drop-in gate performs local model inference in milliseconds
without monitoring API calls.
\end{abstract}

\section{Introduction}

Large language model (LLM) agents combine planning, tool use, and memory to carry out complex tasks, enabling applications ranging from digital assistants to knowledge-intensive decision support \citep{debenedetti2024agentdojo,chen2024agentpoison}. Their ability to interact with external environments also creates security risks: attackers can embed malicious instructions in tool outputs or poison stored knowledge to redirect an agent toward unauthorized actions \citep{debenedetti2024agentdojo,chen2024agentpoison,zhang2025agent}. These risks become particularly challenging when malicious intent is distributed across multiple interactions. Decomposition attacks, for example, split a harmful objective into seemingly benign subtasks that can evade request-level safeguards \citep{yueh2026monitoring}, while memory poisoning allows adversarial information to influence later decisions through retrieval \citep{chen2024agentpoison}. Existing defenses include lightweight sequential monitors that detect harmful intent from cumulative prompt histories \citep{yueh2026monitoring}. Nevertheless, protecting agents throughout extended interactions requires accounting for how attack-relevant information is introduced, retained, and subsequently used during execution.

\begin{figure*}[t]
    \centering
    \includegraphics[width=\textwidth]{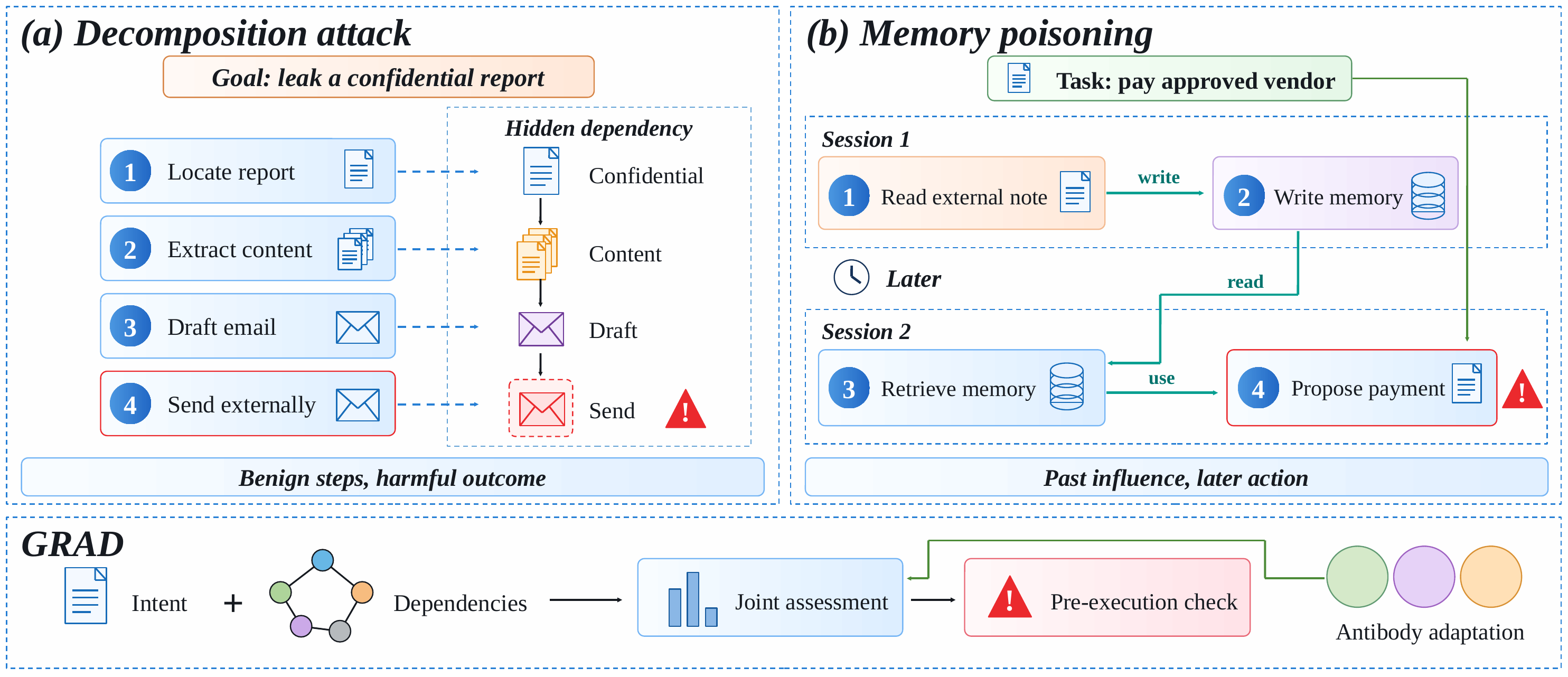}
    \vspace{-4mm}
    \caption{\textbf{Decomposition, memory poisoning, and GRAD.}
    (a) A decomposition attack spreads a confidential-report leak across
    individually ordinary operations, from locating the report to sending
    its contents externally.
    (b) Memory poisoning stores untrusted information in one session and
    reuses it in a later session to influence a proposed payment.
    GRAD jointly assesses operation content and candidate execution
    relations for pre-execution risk detection, while its antibody
    mechanism acquires and reuses task-specific detection experts.}
    \label{fig:threat_scenarios}
    \vspace{-5mm}
\end{figure*}

Figure~\ref{fig:threat_scenarios} illustrates how earlier operations influence a later action. For example, an agent may retrieve external content, store a summary, and later consult it when preparing a tool call. Memory-poisoning studies show that malicious records can influence subsequent behavior through retrieval \citep{chen2024agentpoison,dong2025memory}; the challenge is to assess the content being reused together with the context of its use. Sequential monitors accumulate conversational evidence \citep{yueh2026monitoring}. We study how to organize this evidence across operations. We investigate typed execution graphs as a representation that combines operation content with candidate relations inferred from event order, operation types, and textual cues. These relations provide paths for aggregating evidence across operations. This design targets decomposition attacks, whose harmful intent is distributed across interactions, while the operation-text channel covers locally visible cues such as poisoned tool content.

Beyond combining evidence within an execution, a detector must adapt to
changing tasks while retaining useful capabilities learned from earlier
ones. This setting includes both new instances of familiar attack
strategies and attacks from previously unseen families. The
corresponding learning objectives are task adaptation and retention;
a separate family-isolated stream probes transfer to held-out attack
families. Updating a shared detector can improve a new task while
altering its performance on earlier tasks. This tension has been studied
in continual adversarial robustness~\citep{dai2025adapting} and lifelong
safety alignment~\citep{wang2025lifelong}.

To address these challenges, we propose \textbf{GRAD}, a framework for graph-based attack detection and continual adaptation in LLM agents. Its \textbf{Semantic--Structural Understanding Module (SSUM)} combines operation content and event attributes with candidate relations in a typed execution graph. Building on this detector, \textbf{Continual Antibody Adaptation (CAA)} constructs a growing collection of detector specialists, termed antibodies: compact, task-specific modules that can be added to a running collection and reused as new tasks arrive. We instantiate the standalone detector as a drop-in pre-execution gate that performs local model inference in milliseconds and supports blocking and replanning in closed-loop environments.

Our contributions are threefold:
\begin{itemize}
    \item We propose
    GRAD, an execution-graph framework whose SSUM encoder integrates
    operation semantics, event attributes, and typed candidate relations
    for agent attack detection. Across Decomp, MCP, and MRT, GRAD
    outperforms locally adapted Agent-Sentry and NeuroTaint baselines,
    achieving 97.19, 77.59, and 86.60 AUROC, respectively. We
    additionally report Sequential Monitor results under native Decomp
    and adapted MCP prefix protocols; matched decision-time comparisons
    appear in Section~\ref{sec:overall_comparison}.

    \item We develop CAA, which combines encoder adaptation, frozen
    snapshots, and graph-signature routing into a growing expert
    collection. On the cross-dataset
    stream, CAA improves final accuracy over Replay by 13.08 points
    and reduces forgetting from 0.0629 to 0.0079; equal family/base-case
    weighting within each task preserves a 9.05-point accuracy gain.

    \item We instantiate GRAD as a compact drop-in pre-execution gate
    that performs local model inference in milliseconds and supports blocking
    and replanning without monitoring API calls. Benchmark
    comparisons, encoder ablations, and exploratory closed-loop
    evaluation characterize its detection performance and
    security--utility tradeoffs.
\end{itemize}

\section{Related Work}




\subsection{Memory Poisoning and Decomposition Attacks}

AgentPoison poisons memory or knowledge bases and optimizes retrieval
triggers~\cite{chen2024agentpoison}; PoisonedRAG corrupts retrieved
knowledge~\cite{zou2025poisonedrag}; MINJA induces malicious memory
records through query-only interaction~\cite{dong2025memory}.
Decomposition attacks instead distribute harmful intent across inputs:
DrAttack decomposes and reconstructs prompts~\cite{li2024drattack},
while DecomposedHarm studies stateful subtasks~\cite{yueh2026monitoring}.
These settings motivate assessing operation content together with
earlier actions.

\begin{figure*}[t]
    \centering
    \includegraphics[width=\textwidth]{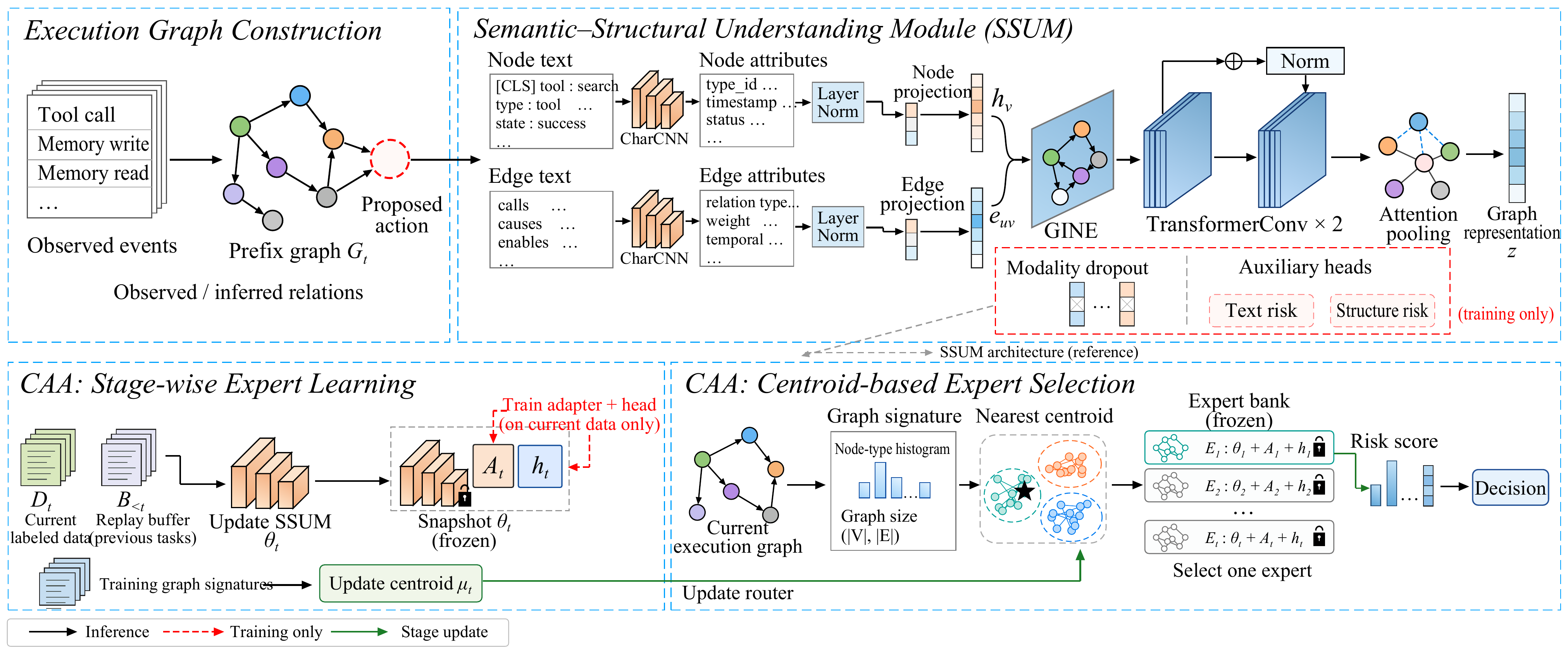}
    \vspace{-4mm}
    \caption{\textbf{GRAD method overview.} Observed operations form an
    execution graph whose text, attributes, and candidate relations are
    encoded by SSUM. CAA updates SSUM on current data and replay, freezes
    stage-specific snapshots, and selects one expert through centroid
    routing for graph-level risk prediction.}
    \label{fig:pipeline}
    \vspace{-5mm}
\end{figure*}
\FloatBarrier

\subsection{Defenses for LLM Agents}

StruQ separates input channels through structured queries and
training~\cite{chen2025struq}; SecAlign uses preference
optimization~\cite{chen2025secalign}. CaMeL enforces tool-use policies
through trusted control/data flows and capabilities~\cite{debenedetti2025defeating},
whereas IPIGuard separates planning from untrusted interactions through
a tool dependency graph~\cite{an2025ipiguard}.
WildGuard classifies harmfulness and refusal~\cite{han2024wildguard},
and sequential monitors accumulate sub-prompt histories~\cite{yueh2026monitoring}.
GUARDIAN models temporal communication graphs~\cite{zhou2025guardian}.
GRAD instead performs supervised risk prediction over operation graphs,
aggregating content through candidate execution relations.

\subsection{Continual Learning for AI Safety}
Regularized continual robust training preserves robustness against
successive adversaries~\cite{dai2025adapting}, while Lifelong Safety Alignment
iteratively updates language-model defenders~\cite{wang2025lifelong}.
Dark Experience Replay retains earlier behavior through
rehearsal~\cite{buzzega2020dark}, and Learning to Prompt selects from a
learned prompt pool with a frozen backbone~\cite{wang2022learning}.
GUARDIAN incorporates incremental training into graph-based
monitoring~\cite{zhou2025guardian}. CAA studies the complementary design
of acquiring frozen detector snapshots and routing among them to support
adaptation and retention across agent tasks.

\section{Method}
\label{sec:method}

\subsection{Problem Formulation and Scope}
\label{sec:problem_formulation}
\label{sec:threat_model}

We study graph-level detection of memory poisoning and decomposition
attacks, followed by adaptation across a stream of labeled tasks.
At stage $r$, $\mathcal{D}_r=\{(G_i,y_i)\}_{i=1}^{N_r}$ contains
operation graphs with graph-level labels $y_i\in\{0,1\}$ for benign and
attack behavior; each $G_i$ organizes the operation content and
candidate relations available to the monitor.

\noindent\textbf{Attacker and observation boundary.}
The attacker controls subrequest content and ordering or contaminates
memory through record insertion.
The monitoring infrastructure, detector parameters, and labeled
adaptation data are trusted. Access and budgets are experiment-specific;
Appendix~H.1 formalizes the attack settings.
The monitor observes recorded operation content, event attributes,
and the proposed action; graph construction uses these inputs.

\noindent\textbf{Detection objective.}
A detector $f_\Theta(G)\in[0,1]$ assigns graph-level risk. Its base
supervised objective on a labeled set $\mathcal{D}$ is
\begin{equation}
    \mathcal{L}_{\mathrm{det}}(\Theta;\mathcal{D})
    =\frac{1}{|\mathcal{D}|}\sum_{(G,y)\in\mathcal{D}}
    w(G,y)\,\ell_{\mathrm{BCE}}\bigl(y,f_\Theta(G)\bigr),
    \label{eq:detection_objective}
\end{equation}
where $w(G,y)$ is the protocol-specified training weight and
$\ell_{\mathrm{BCE}}(y,p)=-y\log p-(1-y)\log(1-p)$.
Continual adaptation seeks to improve detection on incoming tasks
while retaining performance on earlier tasks. A task stage specifies
a data distribution and may contain attack families encountered earlier;
family-isolated evaluation separately measures transfer to unseen families.

For runtime use, $G_t=\Phi(\mathcal{E}_{<t},a_t)$ contains the observed
history $\mathcal{E}_{<t}$ and the proposed action $a_t$, including
its description and arguments available before execution. A score
$f_\Theta(G_t)\geq\tau$ triggers intervention, with $\tau$ selected
on development data. Offline evaluation measures graph classification,
while closed-loop evaluation measures attack success and legitimate
task completion after intervention.

\subsection{Overview of GRAD}
\label{sec:overview}

GRAD first learns a graph representation for risk prediction through
SSUM (Section~\ref{sec:ssum}), then expands a collection of detector
specialists through CAA (Section~\ref{sec:caa}). Each new task adds an expert and
a routing signature; inference selects an expert from the resulting
collection, as shown in Figure~\ref{fig:pipeline}. For closed-loop intervention, a separately trained SSUM
detector scores proposed actions before execution. Local graph
construction and neural inference produce the gate decision; the agent
executes or replans according to that decision. The full procedure is
provided in Appendix~H.4.

\subsection{Semantic--Structural Understanding Module}
\label{sec:ssum}

SSUM encodes operation content together with relations between events
to predict graph-level risk.

\noindent\textbf{Graph and input encoding.}
For an execution graph $G=(V,E)$, nodes follow recorded event order.
We connect adjacent events and add typed candidate edges from operation
types and textual cues. Each edge instance $e=(u,v,r)$ carries a
relation type $r$, heuristic weight $s_e$, and description $d_e$.
Node features comprise text $c_v$ and an attribute vector
$\mathbf{x}_v=[\mathbf{p}_v\Vert\rho_v\Vert\mathbf{k}(c_v)]$
containing permissions, normalized position, and lexical indicators.
Separate character CNNs $C_n,C_e$ encode up to 512 node characters
and 160 edge characters, respectively. Normalized
feature blocks are concatenated and projected:
\begin{equation}
\begin{aligned}
\mathbf{h}_v^{(0)}&=W_n[\operatorname{LN}_n(\mathbf{x}_v)
 \Vert\operatorname{LN}_{nt}(C_n(c_v))]+b_n,\\
\mathbf{e}_e&=W_e[\operatorname{LN}_e([\operatorname{onehot}(r)\Vert s_e])
 \Vert\operatorname{LN}_{et}(C_e(d_e))]+b_e.
\end{aligned}
\label{eq:joint_encoding}
\end{equation}
Layer normalization scales each input block before joint encoding.
Graph-construction rules and dimensions are given in
Appendix~H.3.

\noindent\textbf{Message passing and risk prediction.}
One GINE layer followed by two TransformerConv layers aggregates
edge-conditioned messages. For layer $\ell$, let $\mathcal{M}_\ell$
denote its aggregation operator:
\begin{equation}
\begin{aligned}
\mathbf{z}_v^{(\ell)}&=\mathcal{M}_\ell\!\left(
 \mathbf{h}_v^{(\ell-1)},
 \{(\mathbf{h}_u^{(\ell-1)},\mathbf{e}_e):e=(u,v,r)\in E\}\right),\\
\mathbf{h}_v^{(\ell)}&=\mathbf{h}_v^{(\ell-1)}+
 \operatorname{Dropout}\!\left(\operatorname{GELU}
 (\operatorname{BN}_\ell(\mathbf{z}_v^{(\ell)}))\right),
 \quad \ell=1,2,3.
\end{aligned}
\label{eq:message_passing}
\end{equation}
Learned attention pools the final node states into a graph embedding
$\mathbf{g}$, which a classification head $H_0$ maps to risk:
\begin{equation}
\alpha_v=\sigma(\mathbf{w}_a^\top\mathbf{h}_v^{(3)}+b_a),\qquad
\mathbf{g}=\frac{\sum_v\alpha_v\mathbf{h}_v^{(3)}}
 {\max(\sum_v\alpha_v,\varepsilon)},\qquad
p_{\mathrm{SSUM}}(G)=\sigma(H_0(\mathbf{g})).
\label{eq:graph_prediction}
\end{equation}

\noindent\textbf{Training.}
We apply graph-wise modality dropout to the node inputs, independently
masking permission/position features and lexical/text features with
probability $p_m=0.25$. Two auxiliary classifiers supervise pooled
node-text and permission/position representations
(Appendix~H.3). Joint training uses
\begin{equation}
\mathcal{L}_{\mathrm{SSUM}}=
\mathbb{E}_{m}\!\left[\mathcal{L}_{\mathrm{det}}^{(m)}+
\lambda_{\mathrm{aux}}(\mathcal{L}_{\mathrm{text}}^{(m)}+
\mathcal{L}_{\mathrm{attr}}^{(m)})\right],
\qquad \lambda_{\mathrm{aux}}=0.1,
\label{eq:ssum_training}
\end{equation}
where $m$ denotes the sampled masks, $\mathcal{L}_{\mathrm{det}}^{(m)}$
is Eq.~\ref{eq:detection_objective} on masked inputs, and the auxiliary
terms are binary cross-entropy losses with weights specified in
Appendix~H.3. At inference, all input blocks
are active and the joint head produces the risk score. This encoder
provides the representation on which CAA builds detector specialists.

\subsection{Continual Antibody Adaptation}
\label{sec:caa}

CAA maintains a sequence of encoder snapshots and task-specific
predictors. At stage $r$, a working backbone is updated with current
examples and a bounded buffer of historical examples with cached
probabilities. Its selected parameters are copied to a frozen snapshot
$\theta_r^\star$, on which the new antibody is trained. Previous
antibodies retain their encoder snapshots. The replay objective and
snapshot procedure are detailed in Appendix~L.

\noindent\textbf{Expert acquisition.}
With frozen snapshot parameters $\theta_r^\star$, expert $r$ consists of a residual
adapter $A_r$ and classification head $H_r$:
\begin{equation}
A_r(\mathbf{z})=\mathbf{z}+\mathbf{U}_r
 \operatorname{GELU}(\mathbf{V}_r\mathbf{z}),\qquad
p_r(G)=\sigma(H_r(\mathbf{g}_r(G))).
\label{eq:antibody_adapter}
\end{equation}
The adapter transforms the final message-passing output
$\mathbf{z}_v^{(3)}$ in Eq.~\ref{eq:message_passing}; the snapshot normalization
and pooling then produce $\mathbf{g}_r(G)$.
Here $\mathbf{V}_r\in\mathbb{R}^{b\times d}$ and
$\mathbf{U}_r\in\mathbb{R}^{d\times b}$ are bottleneck projections.
Their configuration and placement are detailed in
Appendix~H.5.
Training optimizes the new parameters
$\phi_r=\{\mathbf{U}_r,\mathbf{V}_r,H_r\}$ on task $r$:
\begin{equation}
\min_{\phi_r}\frac{1}{|\mathcal{D}_r|}
 \sum_{(G,y)\in\mathcal{D}_r}\ell_{\mathrm{BCE}}(y,p_r(G)),
\qquad \theta_r=\theta_r^\star,\quad
 (\theta_j,\phi_j)=(\theta_j^\star,\phi_j^\star)\ (j<r).
\label{eq:expert_learning}
\end{equation}

\noindent\textbf{Expert reuse.}
Each task stores the mean training-graph signature
$\boldsymbol{\mu}_r$. The signature $\mathbf{s}(G)$ summarizes node
 types and graph size (Appendix~H.5). At stage $r$,
nearest-centroid routing selects an expert and returns its prediction:
\begin{equation}
j^\star(G)=\arg\min_{j\leq r}
 \|\mathbf{s}(G)-\boldsymbol{\mu}_j\|_2,\qquad
p_{\mathrm{CAA}}(G)=p_{j^\star(G)}(G).
\label{eq:antibody_routing}
\end{equation}
The expert collection thus grows through task-specific updates and
uses the input graph to select a predictor at inference.

\section{Experiments}
\label{sec:exp}

\subsection{Experimental Setup}
\label{sec:expset}

\noindent\textbf{Datasets.}
Decomp, derived from DecomposedHarm~\citep{yueh2026monitoring}, evaluates
stateful decomposition attacks. MCP and MRT adapt
MCP-SafetyBench~\citep{zong2026mcp} and MT-AgentRisk~\citep{li2026unsafer}
for tool-use risk detection. These three datasets provide the final
GRAD-512 comparisons. MemoryGraft~\citep{srivastava2025memorygraft}
and MINJA~\citep{dong2025memory} provide legacy memory-poisoning
diagnostics in Appendix~B, Table~S6. Construction and splits
appear in Appendix~A; AgentDojo diagnostics appear in
Appendix~C.

\noindent\textbf{Comparisons.}
We compare GRAD with locally adapted Agent-Sentry and NeuroTaint
components and additional agent defenses. Appendix~B
details their observation protocols and implementation mappings.
The Text-only and temporal-detector comparison matches observation
horizon, text budget, splits, and training protocol; a cached sequential
monitor supplies a binary operating point. CAA is compared with
sequential fine-tuning, Replay, and alternative routing policies
(Appendix~L).

\noindent\textbf{Metrics.}
Detection is measured by AUROC; binary decisions are evaluated with
true-positive and false-positive rates (TPR/FPR). For a stream of $R$
tasks, let $a_{r,j}$ denote accuracy on task $j$ after stage $r$.
We report final average accuracy and forgetting:
\begin{equation}
\operatorname{ACC}_R=\frac{1}{R}\sum_{j=1}^{R}a_{R,j},\qquad
F_R=\frac{1}{R}\sum_{j=1}^{R}
 \left[\max_{r=j,\ldots,R}a_{r,j}-a_{R,j}\right].
\label{eq:cl_metrics}
\end{equation}
Closed-loop evaluation reports attack success rate (ASR), benign task
completion (BU), legitimate task completion under attack (UA), and
runtime errors. Latency measurements specify their timing scope in
Appendix~G.

\noindent\textbf{Implementation.}
Training uses NVIDIA RTX 3090 GPUs. Thresholds are selected on
development data; seed counts and standard-deviation conventions are
specified with each result. \FinalStreamProtocol.
Additional configurations and aggregation details appear in
Appendices~D--G.

\subsection{Comparison with Existing Defenses}
\label{sec:overall_comparison}

Tables~\ref{tab:detection} and~\ref{tab:binary_checking} compare GRAD
with the evaluated defense implementations on Decomp, MCP, and MRT.
The score-based rows report locally adapted components of Agent-Sentry
and NeuroTaint under our shared observable input; Sequential Monitor
is evaluated under a native Decomp and adapted MCP prefix protocol,
and is additionally compared in the closed loop
(Table~\ref{tab:backbones}). The tables retain each implementation's
observation and threshold protocols; the matched Decomp comparison
below controls these factors.

\par
\begin{wraptable}{r}{\textwidth}
\centering\small
\caption{Detection AUROC (\%). GRAD reports mean and sample standard
deviation across three seeds. Agent-Sentry and NeuroTaint rows are
locally adapted components, not the complete defenses; the Sequential
Monitor uses its declared prefix adaptation, and its MRT entry is
unavailable because that adaptation was not executed
(Appendix~B). Implementation and
observation protocols: Appendices~B.1 and~B.}
\label{tab:detection}
\begin{adjustbox}{max width=\linewidth}
\begin{tabular}{lccc}
\toprule
Method & Decomp & MCP & MRT \\
\midrule
Sequential Monitor & 81.03 & 56.16 & N/A \\
Agent-Sentry L1 & 84.44 & 53.12 & 51.90 \\
NeuroTaint propagation & 72.53 & 50.00 & 50.00 \\
\midrule
\textbf{GRAD} & $\mathbf{97.19\pm0.99}$ & $\mathbf{77.59\pm2.01}$ & $\mathbf{86.60\pm5.13}$ \\
\bottomrule
\end{tabular}
\end{adjustbox}
\end{wraptable}
\WFclear

\begin{wraptable}{r}{\textwidth}
\centering\small
\caption{TPR/FPR (\%) at reported operating points. GRAD uses per-run
development thresholds. Implementation and threshold protocols:
Appendix~B.1.
AP/AN denote all-positive/all-negative predictions; --- denotes
an unavailable operating point.}
\label{tab:binary_checking}
\begin{adjustbox}{max width=\linewidth}
\begin{tabular}{lccc}
\toprule
Method & Decomp & MCP & MRT \\
\midrule
Sequential Monitor & 100.0/36.2 & N/A & N/A \\
Agent-Sentry L1 & 74.1/15.5 & 15.8/7.6 & --- \\
NeuroTaint propagation & 25.9/8.6 & $0.0/0.0^{\mathrm{AN}}$ & $0.0/0.0^{\mathrm{AN}}$ \\
AgentTrust + RiskChain & $0.0/0.0^{\mathrm{AN}}$ & 3.5/0.0 & 20.0/12.0 \\
AgentArmor & 98.9/91.4 & 64.9/69.5 & $100.0/100.0^{\mathrm{AP}}$ \\
AEGIS & 23.0/6.9 & 54.4/54.2 & $100.0/100.0^{\mathrm{AP}}$ \\
\midrule
\textbf{GRAD} & \ResultCell{det512-decomp-tpr}/\ResultCell{det512-decomp-fpr} & \ResultCell{det512-mcp-tpr}/\ResultCell{det512-mcp-fpr} & \ResultCell{det512-mrt-tpr}/\ResultCell{det512-mrt-fpr} \\
\bottomrule
\end{tabular}
\end{adjustbox}
\end{wraptable}
\WFclear

\noindent\textbf{Decision-time comparison.}
We truncate Decomp attacks at the proposed harmful action, retaining
that action and all preceding observations. GRAD, Text-only, and
Temporal receive identical 200-character step strings and are trained
on the same truncated splits with matched budgets and development
selection. The test set contains 232 cases in 58 base-case groups.

\par
\begin{wraptable}{r}{\textwidth}
\centering\small
\caption{Decomp decision-time detection (\%). Learned detectors report
three-seed mean AUROC $\pm$ population SD and mean TPR/FPR at
individually development-selected thresholds. The cached binary monitor
uses its native decision rule.}
\label{tab:w1_decomp}
\begin{tabular}{lrrr}
\toprule
Method & AUROC $\uparrow$ & TPR $\uparrow$ & FPR $\downarrow$ \\
\midrule
\textbf{GRAD} & $\mathbf{97.0\pm1.2}$ & 89.7 & 8.0 \\
Temporal (CharCNN + Transformer) & $91.7\pm0.5$ & 77.6 & 13.2 \\
Text-only (CharCNN) & $86.4\pm1.7$ & 78.2 & 20.1 \\
Length-only control & 77.7 & --- & --- \\
\midrule
Sequential Monitor (binary) & --- & 100.0 & 37.9 \\
\bottomrule
\end{tabular}
\end{wraptable}
\WFclear

On the original split, GRAD achieves 97.0\% AUROC, exceeding Temporal and Text-only by
5.30 and 10.64 percentage points, respectively, as shown in Table~\ref{tab:w1_decomp}.
Paired base-case bootstrap 95\% intervals are [2.0, 9.1] and
[7.2, 14.4] points, both excluding zero. GRAD also achieves
higher TPR and lower FPR than both learned controls at their selected
operating points. The monitor detects all attacks while flagging 37.9\%
of benign cases; an ordinal score derived from its first-trigger
position reaches 91.24\% AUROC on the same truncated cases, still below
GRAD's 97.0\% while sharing the decision-time window, and is a derived
ranking rather than a native continuous output. The length-only AUROC
of 77.7\% motivates retaining trace-length controls. With a 5\% development FPR target, GRAD detects 85.6\% of attacks
at 7.5\% test FPR, versus 54.4/9.2 for Text-only and 67.8/5.2 for
Temporal. Full threshold results, paired statistics, and costs appear in
Appendix~M.
On 47 independently generated base cases, the frozen GRAD--Temporal
AUROC difference is $-0.060$ with a 95\% CI of [$-0.115,-0.0001$];
the original advantage does not reproduce (Appendix~P).

The encoder ablations below use the original Decomp full-trace and
MCP/MRT protocols. Table~\ref{tab:binary_checking} also shows the
operating-point variation: MRT reaches 83.3\% TPR at 58.0\% FPR under
its development-selected thresholds.

\Needspace{14\baselineskip}
\subsection{Encoder Ablations}
\label{sec:ablation}

\noindent\textbf{Encoder comparison.}
Attack evidence can be expressed in operation content or in the
relationships between operations.

\par
\begin{wraptable}{r}{0.62\textwidth}
\centering\footnotesize\setlength{\tabcolsep}{3pt}
\caption{Encoder ablations: AUROC (\%) as mean and sample SD across
three seeds in the final 512-character replication batch.
Protocol details: Appendix~K.}
\label{tab:ablation}
\begin{adjustbox}{max width=\linewidth}
\begin{tabular}{lccc}
\toprule
Variant & Decomp & MCP & MRT \\
\midrule
\textbf{GRAD} & $\mathbf{97.19\pm0.99}$ & $77.59\pm2.01$ & $\mathbf{86.60\pm5.13}$ \\
Graph only & $86.37\pm0.79$ & $67.41\pm0.89$ & $86.07\pm3.95$ \\
Text only & $86.83\pm0.48$ & $\mathbf{79.89\pm2.72}$ & $82.87\pm6.40$ \\
\bottomrule
\end{tabular}
\end{adjustbox}
\end{wraptable}

Explicit malicious instructions can
provide strong textual cues, whereas the risk of reusing information
can depend on its source and subsequent use. Table~\ref{tab:ablation}
compares these input channels in the same final replication batch.
On Decomp, where harmful intent is distributed across operations, GRAD
exceeds text-only and graph-only encoders by approximately 10.4 and
10.8 AUROC points, respectively; neither single channel exceeds 87\%.
On MRT, graph-only attains 86.07\%, close to GRAD's 86.60\% and above
text-only's 82.87\%. The graph-only arm uses event attributes as well
as topology.

\noindent\textbf{Generalization to other attacks.}
MCP evaluates poisoned tool content, outside the settings that
motivate the relational design. GRAD attains 77.59\%, the strongest
score among the defenses in Table~\ref{tab:detection},
close to text-only at 79.89\% and above graph-only at 67.41\%.

\par\WrapTableClear
\subsection{Continual Learning of Antibody Experts}
\label{sec:continual_learning}

We evaluate adaptation and retention across changing task distributions
on the Decomp/MCP/MRT stream under its sample-level splits, with attack
families recurring across training and evaluation.
Snapshot-based CAA is evaluated with five seeds and forward/reverse task orders. Sequential fine-tuning uses current-task examples; Replay and snapshot
updates also use buffered historical examples. All arms share the 512-character
window, architecture, splits, and optimization settings.
Table~\ref{tab:continual} separately reports a family-isolated,
source-balanced mechanism stream that probes transfer to held-out
families. Protocols appear in Appendix~L.

\par
\begin{center}
\small
\captionof{table}{Continual-learning results with a common 512-character backbone within
 each panel. Task-macro accuracy, averaged over five seeds and two orders. Cross-dataset tasks
 are Decomp/MCP/MRT; the mechanism stream is family-isolated.
 Implementation and paired intervals: Appendix~L.}
\label{tab:continual}
\begin{adjustbox}{max width=\linewidth}
\begin{tabular}{lcccc}
\toprule
 & \multicolumn{2}{c}{Cross-dataset} & \multicolumn{2}{c}{Mechanism stream} \\
\cmidrule(lr){2-3}\cmidrule(lr){4-5}
Method & ACC (\%) & $F_R$ & ACC (\%) & $F_R$ \\
\midrule
Sequential FT & 56.43 & 0.1251 & 45.62 & 0.0797 \\
Replay & 64.78 & 0.0629 & 39.21 & 0.0454 \\
\textbf{Snapshot experts + routing} & \textbf{77.86} & \textbf{0.0079} & \textbf{51.11} & \textbf{0.0327} \\
\bottomrule
\end{tabular}
\end{adjustbox}
\end{center}

\noindent\textbf{Cross-task adaptation and retention.}
On the cross-dataset stream, CAA increases final accuracy from
Replay's 64.78\% to 77.86\%, a \textbf{13.08-percentage-point gain}
with a 95\% CI of 9.11--19.10, while reducing forgetting from 0.0629 to
0.0079; the paired difference is -0.0551 with a 95\% CI of -0.1185-- -0.0169.
With equal weights for families/base cases within each task, the
accuracy gain remains \textbf{9.05 points} with a 95\% CI of 6.25--11.91.
The improvement is driven by Decomp and MRT: their gains are 18.33
and 17.09 points under the original endpoint, and 18.38 and 15.49
points under family-equal weighting.

\noindent\textbf{Robustness across weights and resampling.}
The original accuracy gain remains positive under two-way seed
and task-order resampling with a 95\% CI of 8.44--21.15 points.
A sample-pooled family-cluster analysis also yields a positive
interval of 7.72--19.94 points.
After removing the largest families under the size-sensitivity rule,
Decomp and MRT retain gains of 18.66 and 14.86 points.
MCP contributes a smaller 3.82-point sample-level gain; its
family-equal difference is -6.72 points with a 95\% CI of -17.01--1.38.
Appendix~L.1 reports the complete analysis.

\noindent\textbf{Expert selection.}
Centroid routing achieves 77.86\% accuracy on the cross-dataset snapshot
pool, compared with 71.58\% for all-expert averaging and 79.86\% for
task-oracle routing, so routing selection matters more than ensemble
size. Appendix~L reports the routing
comparison and costs. \FinalRoutingResults

\noindent\textbf{Family-isolated boundary.}
Snapshot experts attain 51.11\% accuracy and 0.0327 forgetting,
compared with Replay's 39.21\% and 0.0454. The paired accuracy gain
is 11.91 percentage points with a 95\% CI of 3.95--19.30; the forgetting
difference is -0.0127 with a 95\% CI of -0.0534--0.0315.
The sample-level and family-equal endpoints emphasize different parts
of this stream: equal weighting of label-pure families gives 47.5\%
for snapshot experts and 53.5\% for Replay. The paired difference
is -6.0 points with a 95\% CI of -9.7-- -2.2;
Appendix~L.2 reports the weighting analysis.
The reversal under equal family weighting identifies family composition
as the boundary condition on snapshot gains in this stream; this panel
is reported as a held-out-family boundary rather than uniform transfer
evidence.

\noindent\textbf{Contribution of backbone updates.}
Holding the expert-library design and centroid routing fixed, stage-0
experts attain 68.37\% accuracy on the cross-dataset stream, compared
with 77.86\% for stage snapshots: a paired gain of 9.49 percentage
points with a 95\% CI of 8.34--11.02. On the family-isolated stream, stage-0
experts attain 53.18\% accuracy and 0.0042 forgetting; snapshot updates
increase forgetting by 0.0285 with a 95\% CI of 0.0034--0.0667.
The ablation separates the benefit of acquiring and routing among
experts from the stream-dependent benefit of updating their backbones.

\subsection{Exploratory Transfer and Closed-Loop Evaluation}
\label{sec:backbone_evaluation}

In the offline transfer comparison, the from-scratch gate reaches 94.77\% gate-test
AUROC but 50.31\% on Decomp, the Decomp-trained detector 68.60\% on gate
test and 94.28\% on Decomp, and the joint adaptation 96.08\% and 96.88\%
as reported in Appendix~F, Table~S14.

\par
\begin{wraptable}{r}{\textwidth}
\centering\small
\caption{Closed-loop defense comparison with the transfer-adapted GRAD gate.
Results are percentages; each arm has 20 attack and 20 benign runs per
backbone.}
\label{tab:backbones}
\begin{adjustbox}{max width=\linewidth}
\begin{tabular}{llrrr}
\toprule
Backbone & Defense & ASR$\downarrow$ & BU$\uparrow$ & UA$\uparrow$ \\
\midrule
DeepSeek Flash & None & 35 & 100 & 50 \\
 & Sequential Monitor & 0 & 50 & 25 \\
 & GRAD & 0 & 75 & 80 \\
\midrule
GLM-5.3-Flash & None & 30 & 90 & 45 \\
 & Sequential Monitor & 0 & 50 & 0 \\
 & GRAD & 10 & 40 & 30 \\
\bottomrule
\end{tabular}
\end{adjustbox}
\end{wraptable}
\WFclear

Table~\ref{tab:backbones} evaluates one jointly adapted checkpoint and
threshold across four attack families on both backbones over 240 runs.
On DeepSeek, GRAD and Sequential Monitor both achieve 0\% ASR,
with benign completion of 75\% and 50\%, respectively. On GLM,
GRAD achieves 10\% ASR and 40\% benign completion, versus 0\% and
50\% for the monitor. GRAD's legitimate completion under attack is
80\% on DeepSeek and 30\% on GLM, compared with 25\% and 0\% for
the monitor. These outcomes characterize backbone-dependent
security--utility tradeoffs. Appendix~F reports both
batches and paired statistics.

\subsection{Online Scoring Time}
\label{sec:online_cost}

\par
\begin{wraptable}{r}{0.62\textwidth}
\centering\footnotesize\setlength{\tabcolsep}{3pt}
\caption{Logged gate-forward and monitor-call latency in the matched
closed-loop batch (ms). The gate timer excludes graph preparation;
the monitor timer covers the hosted judgment call.}
\label{tab:online_cost}
\begin{adjustbox}{max width=\linewidth}
\begin{tabular}{llrr}
\toprule
Backbone & Defense & p50 & p95 \\
\midrule
DeepSeek & Monitor & 603.3 & 796.1 \\
 & GRAD & 8.4 & 33.2 \\
\midrule
GLM & Monitor & 619.4 & 945.5 \\
 & GRAD & 8.3 & 32.1 \\
\bottomrule
\end{tabular}
\end{adjustbox}
\end{wraptable}

Table~\ref{tab:online_cost} reports gate-forward and monitor-call latency.
GRAD requires no monitoring API calls. Median model-forward times are
8.4/8.3\,ms on DeepSeek/GLM traces, versus 603.3/619.4\,ms for monitor
calls. Graph preparation is outside the gate timer; measurement details
appear in Appendix~K.4.

\par\WrapTableClear
\section{Discussion}
\label{sec:discussion}

\noindent\textbf{Benchmark detection.}
Decomposition attacks show the largest joint-encoder improvement over
both single-channel arms; the gain is not attributable to a single
channel there, as both single-channel arms remain below 87\% AUROC.
MRT graph-only and joint encoding perform
similarly. On MCP, text-only encoding attains 79.89\% AUROC,
compared with 77.59\% for GRAD.

\noindent\textbf{Adaptation and retention.}
CAA improves cross-dataset accuracy and retention over Replay, with
Decomp and MRT providing the largest gains. These results concern
sequential task contexts with recurring attack families; the separate
family-isolated stream probes held-out-family behavior.
Appendix~L separates expert selection and backbone
updates.

\section{Limitations and Conclusion}
\label{sec:conclusion}

\noindent\textbf{Limitations.}
GRAD depends on observable execution records; incomplete logging or
out-of-context information can limit detection. New environments require
labeled data and mappings to graph attributes and candidate relations.
CAA also incurs growing snapshot storage, and held-out-family gains
are endpoint-dependent. Future work will address coverage, label
efficiency, and expert consolidation.

\noindent\textbf{Conclusion.}
GRAD combines execution-graph risk detection with an extensible snapshot
expert library. It achieves 97.0\% AUROC on the original matched Decomp evaluation
versus 91.7\% for the temporal detector, and CAA improves
cross-dataset accuracy over Replay by 13.08 points
and by 9.05 points with equal family/base-case weights within each task. Closed-loop
evaluation characterizes the security--utility tradeoffs of local
blocking and replanning. CAA adds experts on updated backbone snapshots while retaining
previously acquired experts.

\clearpage
\subsection*{AI use statement}
Generative AI tools assisted with writing and polishing, literature
retrieval and discovery, research and experimental execution, manuscript
drafting, and synthetic dataset generation. Appendix~N details these uses and author
responsibility for the final manuscript, references, datasets, and results.

\subsection*{Ethics statement}
This research examines prompt injection and memory poisoning to improve
the security of tool-using agents. The controlled closed-loop evaluations
use simulated tool environments. Actions such as payments and external
transfers in these environments represent simulated outcomes. Attack
payloads and traces have dual-use implications: they can support defensive
evaluation but may also facilitate misuse. Any sharing of evaluation
artifacts must respect source-data permissions and protect credentials,
personal information, and sensitive execution content. Reported results
concern the evaluated environments and do not constitute authorization
to test attacks against external systems.

\subsection*{Reproducibility statement}
Section~\ref{sec:method} defines the detector and adaptation mechanisms,
and Section~\ref{sec:expset} describes the evaluation setup and metrics.
The appendix documents dataset transformations, evaluated baseline
implementations, input windows, training settings, seeds, task orders,
threshold selection, and aggregation procedures. Experiment records
include configuration and split information, per-run outputs, and,
where recorded, code and checkpoint hashes. Static detection,
continual-learning, and closed-loop results are identified by their
respective protocols. Appendix~G details the
artifact provenance and timing definitions; the extended tables provide
confusion counts, seed-level results, and paired comparisons for checking
the reported aggregates.


\bibliography{iclr2027_conference}

@article{debenedetti2024agentdojo,
  title={Agentdojo: A dynamic environment to evaluate prompt injection attacks and defenses for llm agents},
  author={Debenedetti, Edoardo and Zhang, Jie and Balunovic, Mislav and Beurer-Kellner, Luca and Fischer, Marc and Tram{\`e}r, Florian},
  journal={Advances in neural information processing systems},
  volume={37},
  pages={82895--82920},
  year={2024}
}

@article{chen2024agentpoison,
  title={Agentpoison: Red-teaming llm agents via poisoning memory or knowledge bases},
  author={Chen, Zhaorun and Xiang, Zhen and Xiao, Chaowei and Song, Dawn and Li, Bo},
  journal={Advances in Neural Information Processing Systems},
  volume={37},
  pages={130185--130213},
  year={2024}
}

@inproceedings{zhang2025agent,
  title={Agent security bench (asb): Formalizing and benchmarking attacks and defenses in llm-based agents},
  author={Zhang, Hanrong and Huang, Jingyuan and Mei, Kai and Yao, Yifei and Wang, Zhenting and Zhan, Chenlu and Wang, Hongwei and Zhang, Yongfeng},
  booktitle={International Conference on Learning Representations},
  volume={2025},
  pages={35331--35366},
  year={2025}
}

@inproceedings{yueh2026monitoring,
  title={Monitoring decomposition attacks with lightweight sequential monitors},
  author={Yueh-Han, Chen and Joshi, Nitish and Chen, Yulin and Andriushchenko, Maksym and Angell, Rico and He, He},
  booktitle={The Fourteenth International Conference on Learning Representations},
  year={2026}
}

@inproceedings{dong2025memory,
 author = {Dong, Shen and Xu, Shaochen and He, Pengfei and Li, Yige and Tang, Jiliang and Liu, Tianming and Liu, Hui and Xiang, Zhen},
 booktitle = {Advances in Neural Information Processing Systems},
 doi = {10.52202/085713-1554},
 editor = {D. Belgrave and C. Zhang and H. Lin and R. Pascanu and P. Koniusz and M. Ghassemi and N. Chen},
 pages = {46697--46731},
 publisher = {Curran Associates, Inc.},
 title = {{Memory Injection Attacks on LLM Agents via Query-Only Interaction}},
 url = {https://proceedings.neurips.cc/paper_files/paper/2025/file/42a97bbd9844d2bf68596730af80bcdf-Paper-Conference.pdf},
 volume = {38, Main Conference},
 year = {2025}
}

@inproceedings{zou2025poisonedrag,
  title={{PoisonedRAG}: Knowledge corruption attacks to {Retrieval-Augmented} generation of large language models},
  author={Zou, Wei and Geng, Runpeng and Wang, Binghui and Jia, Jinyuan},
  booktitle={34th USENIX Security Symposium (USENIX Security 25)},
  pages={3827--3844},
  year={2025}
}

@inproceedings{li2024drattack,
  title={DrAttack: Prompt decomposition and reconstruction makes powerful LLMs jailbreakers},
  author={Li, Xirui and Wang, Ruochen and Cheng, Minhao and Zhou, Tianyi and Hsieh, Cho-Jui},
  booktitle={Findings of the Association for Computational Linguistics: EMNLP 2024},
  pages={13891--13913},
  year={2024}
}

@inproceedings{chen2025struq,
  title={{StruQ}: Defending against prompt injection with structured queries},
  author={Chen, Sizhe and Piet, Julien and Sitawarin, Chawin and Wagner, David},
  booktitle={34th USENIX Security Symposium (USENIX Security 25)},
  pages={2383--2400},
  year={2025}
}

@inproceedings{chen2025secalign,
  title={Secalign: Defending against prompt injection with preference optimization},
  author={Chen, Sizhe and Zharmagambetov, Arman and Mahloujifar, Saeed and Chaudhuri, Kamalika and Wagner, David and Guo, Chuan},
  booktitle={Proceedings of the 2025 ACM SIGSAC Conference on Computer and Communications Security},
  pages={2833--2847},
  year={2025}
}

@article{han2024wildguard,
  title={Wildguard: Open one-stop moderation tools for safety risks, jailbreaks, and refusals of llms},
  author={Han, Seungju and Rao, Kavel and Ettinger, Allyson and Jiang, Liwei and Lin, Bill Yuchen and Lambert, Nathan and Choi, Yejin and Dziri, Nouha},
  journal={Advances in neural information processing systems},
  volume={37},
  pages={8093--8131},
  year={2024}
}

@article{debenedetti2025defeating,
  title={Defeating prompt injections by design},
  author={Debenedetti, Edoardo and Shumailov, Ilia and Fan, Tianqi and Hayes, Jamie and Carlini, Nicholas and Fabian, Daniel and Kern, Christoph and Shi, Chongyang and Terzis, Andreas and Tram{\`e}r, Florian},
  journal={arXiv preprint arXiv:2503.18813},
  year={2025}
}

@inproceedings{zhou2025guardian,
 author = {Zhou, Jialong and Wang, Lichao and Yang, Xiao},
 booktitle = {Advances in Neural Information Processing Systems},
 doi = {10.52202/085713-0272},
 editor = {D. Belgrave and C. Zhang and H. Lin and R. Pascanu and P. Koniusz and M. Ghassemi and N. Chen},
 pages = {7973--8001},
 publisher = {Curran Associates, Inc.},
 title = {{GUARDIAN: Safeguarding LLM Multi-Agent Collaborations with Temporal Graph Modeling}},
 url = {https://proceedings.neurips.cc/paper_files/paper/2025/file/0bc795afae289ed465a65a3b4b1f4eb7-Paper-Conference.pdf},
 volume = {38, Main Conference},
 year = {2025}
}

@article{dai2025adapting,
  title={Adapting to evolving adversaries with regularized continual robust training},
  author={Dai, Sihui and Cianfarani, Christian and Bhagoji, Arjun and Sehwag, Vikash and Mittal, Prateek},
  journal={arXiv preprint arXiv:2502.04248},
  year={2025}
}

@inproceedings{wang2025lifelong,
 author = {Wang, Haoyu and Zhao, Yifei and Qin, Zeyu and Du, Chao and Lin, Min and Wang, Xueqian and Pang, Tianyu},
 booktitle = {Advances in Neural Information Processing Systems},
 doi = {10.52202/085713-0670},
 editor = {D. Belgrave and C. Zhang and H. Lin and R. Pascanu and P. Koniusz and M. Ghassemi and N. Chen},
 pages = {19908--19938},
 publisher = {Curran Associates, Inc.},
 title = {{Lifelong Safety Alignment for Language Models}},
 url = {https://proceedings.neurips.cc/paper_files/paper/2025/file/1ca2f4528792b31eab7a3e7f6a1c130b-Paper-Conference.pdf},
 volume = {38, Main Conference},
 year = {2025}
}

@article{buzzega2020dark,
  title={Dark experience for general continual learning: a strong, simple baseline},
  author={Buzzega, Pietro and Boschini, Matteo and Porrello, Angelo and Abati, Davide and Calderara, Simone},
  journal={Advances in neural information processing systems},
  volume={33},
  pages={15920--15930},
  year={2020}
}

@inproceedings{wang2022learning,
  title={Learning to prompt for continual learning},
  author={Wang, Zifeng and Zhang, Zizhao and Lee, Chen-Yu and Zhang, Han and Sun, Ruoxi and Ren, Xiaoqi and Su, Guolong and Perot, Vincent and Dy, Jennifer and Pfister, Tomas},
  booktitle={2022 IEEE/CVF conference on computer vision and pattern recognition (CVPR)},
  pages={139--149},
  year={2022},
  organization={IEEE}
}

@article{srivastava2025memorygraft,
  title={MemoryGraft: Persistent compromise of LLM agents via poisoned experience retrieval},
  author={Srivastava, Saksham Sahai and He, Haoyu},
  journal={arXiv preprint arXiv:2512.16962},
  year={2025}
}

@article{li2026unsafer,
  title={Unsafer in many turns: Benchmarking and defending multi-turn safety risks in tool-using agents},
  author={Li, Xu and Yu, Simon and Pan, Minzhou and Sun, Yiyou and Li, Bo and Song, Dawn and Lin, Xue and Shi, Weiyan},
  journal={arXiv preprint arXiv:2602.13379},
  year={2026}
}

@inproceedings{zong2026mcp,
  title={MCP-SafetyBench: A benchmark for safety evaluation of large language models with real-world MCP servers},
  author={Zong, Xuanjun and Shen, Zhiqi and Wang, Lei and Lan, Yunshi and Yang, Chao},
  booktitle={International Conference on Learning Representations},
  volume={2026},
  pages={130709--130731},
  year={2026}
}

@inproceedings{an2025ipiguard,
  title={IPIGuard: A novel tool dependency graph-based defense against indirect prompt injection in LLM agents},
  author={An, Hengyu and Zhang, Jinghuai and Du, Tianyu and Zhou, Chunyi and Li, Qingming and Lin, Tao and Ji, Shouling},
  booktitle={Proceedings of the 2025 Conference on Empirical Methods in Natural Language Processing},
  pages={1023--1039},
  year={2025}
}
\bibliographystyle{iclr2027_conference}

\clearpage
\appendix

\section{Datasets and Evaluation Protocols}
\label{app:data}
The primary test comparisons use Decomp (174 positive/58 negative),
MCP (114/118), and MRT (50/50) evaluation units per seed, with three
seeds. Fold counts are one, five, and five, respectively. Historical
MG and MINJA comparisons use 600/400 and 75/115 units; the derived
InjecAgent, ASB, and AgentHarm baseline panels use 372/372, 50/50,
and 52/52. These are evaluation counts, not total corpus sizes.
The gate split has 172 training, 42 development, and 59 test graphs.
AgentDojo P1/R1 diagnostics are described below.
\subsection{Data Conversion and Leakage Audits}
MG/MINJA retain template/category confounds, and the derived
InjecAgent/ASB panels contain shortcut cues. They are historical diagnostic
panels. The primary Decomp/MCP/MRT experiments use trajectories
derived or adapted from the benchmarks identified in
Section~\ref{sec:expset}. SSUM encodes their operation content and
reconstructs candidate edges from event order, operation types, and
textual cues.
Family isolation for the continual-learning panels is described
in their respective protocols.
\subsection{Model Selection and Calibration Timeline}
The static G3 and CAA replications reuse the previously evaluated
splits. The 512-character configuration was frozen before the 132-run
four-arm static replication (Appendix~\ref{app:window512}); operating
points use stored validation thresholds. CAA selects checkpoints by
development AUROC on its task streams. Its DMM stream evaluates
sequential task acquisition and retention under sample-level splits;
family definitions and train--test overlap appear in
Appendix~\ref{app:g3_endpoint_recheck}. T3 deployment selection uses
gate-development and gate-test AUROC, and its closed-loop evaluation
reuses a fixed case panel across batches (Appendix~\ref{app:runtime}).

\section{Baseline Implementations and Additional Results}
\label{app:baselines}
The static score and binary-checker tables use the component mappings
in Appendix~\ref{app:detection_versions}. Agent-Sentry is its adapted L1
component, NeuroTaint is propagation without counterfactual evaluation,
AgentArmor uses an assumed registry, AgentTrust/RiskChain uses observable
rules, and AEGIS uses the official classifier with text fallback. The
comparison is component-to-component: each defense is represented by
the component evaluable under our shared observable input, and
runtime-dependent stages are documented below.
The native Sequential Monitor Decomp result uses cumulative prefixes
up to the harmful deadline, versus full benign histories. Its adapted
non-Decomp scores remain in the historical table, with the MCP value
repeated in Table~\ref{tab:detection} under its adapted cumulative-prefix protocol. The MRT adaptation was not executed. NeuroTaint lacks the
neutralized full-context traces required by its counterfactual stage.
A full three-layer local adaptation of Agent-Sentry (L1+L2+L3,
following the published method description) was additionally evaluated
on the static panels; it attains ABR 29.3\%/27.2\%/2.7\%/15.8\% and
Util 82.8\%/86.8\%/90.4\%/92.4\% on Decomp/MG/MINJA/MCP, versus the
94--96\% ABR reported on its native benchmarks. Reproduction gaps:
our graph schema lacks per-argument provenance metadata (approximated
from graph structure), the L3 judge model is substituted, and
action-level attack labels are proxies; MRT is not covered because
its flat tool-call traces do not define the retrieval/action role
partition. This locally adapted full pipeline does not transfer to
our data form, so the reported comparisons use the L1 component.
AuthGraph online results below are separate from static classification.

\subsection{Detection Result Versions and Aggregation}
\label{app:detection_versions}
The historical G3\_p25 results use 33 existing runs: three Decomp
runs, 15 MCP runs, and 15 MRT runs, with seeds 211--213. AUROC is
recomputed from per-sample ranks, weighted by $n_+n_-$ across folds
within a seed, and summarized with the sample standard deviation across
seeds. Earlier G3 summaries used population standard deviations.

\par
\begin{wraptable}{r}{\textwidth}
\centering\small
\caption{Historical GRAD-192 encoder run batches: AUROC (\%), mean and sample SD.
The paired MRT entry reuses the archived runs.}
\label{tab:historical_encoder_batches}
\begin{adjustbox}{max width=\linewidth}
\begin{tabular}{lccc}
\toprule
Batch & Decomp & MCP & MRT \\
\midrule
Original, archived & $95.16\pm0.63$ & $62.17\pm1.73$ & $74.00\pm8.59$ \\
Original, paired & $95.75\pm1.72$ & $\mathbf{64.10\pm2.28}$ & $74.00\pm8.59$ \\
G3\_p25 & $\mathbf{96.63\pm1.41}$ & $63.80\pm2.71$ & $\mathbf{87.30\pm4.94}$ \\
\bottomrule
\end{tabular}
\end{adjustbox}
\end{wraptable}
\WFclear

The alignment audit found common split paths, preprocessing and training
settings across original batches; the recorded trainer code difference
concerns output handling. It attributes the batch differences to repeat-run
variation. Recomputed archived MCP AUROC is 62.17\%, compared with
62.10\% from stored trainer summaries. Historical values below retain
their original aggregation for provenance. The original unified
CausalCharGNN batch and the later G3 component batch are distinct.

Historical GRAD-192 thresholds are selected per run on development data with an FPR
constraint of 10\%. Pooled test $(TP,FP,TN,FN)$ counts are
$(481,24,150,41)$ for Decomp, $(126,61,293,216)$ for MCP, and
$(134,86,64,16)$ for MRT. These counts include repeated evaluations
across seeds. MRT uses 32 calibration graphs and 20 admission graphs
per fold and exhibits poor transfer of the calibration operating point.

Baseline results retain their source protocols. Sequential Monitor's
Decomp observations end at the annotated harm deadline for harmful
trajectories; GRAD scores full graphs. Its non-Decomp adaptations are
retained in the historical extended tables rather than treated as native
pipeline results. Agent-Sentry uses an adapted L1 component; NeuroTaint
uses propagation without the counterfactual stage. AgentArmor uses an
assumed registry. Agent-Sentry MRT has stored scores but no recorded native or
development threshold for its operating point. AuthGraph has an 18-run online AgentDojo evaluation
(Table~\ref{tab:authgraph_online}); static classification is unavailable.

Final G3 coverage comprises Decomp, MCP, and MRT. The extended
MG, MINJA, InjecAgent, ASB, and AgentHarm results below use the
legacy unified encoder, with version and implementation annotations.


\par
\begin{wraptable}{r}{\textwidth}
\centering
\caption{
Historical extended detection results: AUROC (\%, $\uparrow$).
Superscripts identify evaluated implementations.
$^\dagger$ marks diagnostic datasets with unresolved
confounds or shortcut evidence.
$^\ddagger$ marks results requiring observation-window
alignment before direct comparison.
``--'' denotes an unavailable result.
}
\label{tab:legacy_detection}

\setlength{\tabcolsep}{5pt}
\renewcommand{\arraystretch}{1.12}

\resizebox{\linewidth}{!}{%
\begin{tabular}{lcccccccc}
    \toprule
    \multirow{2}{*}{\textbf{Method}}
    & \multicolumn{3}{c}{\textbf{Primary Detection Tasks}}
    & \multicolumn{2}{c}{\textbf{Memory Diagnostics}}
    & \multicolumn{3}{c}{\textbf{Additional Derived Tasks}} \\
    \cmidrule(lr){2-4}
    \cmidrule(lr){5-6}
    \cmidrule(lr){7-9}

    & \textbf{Decomp}
    & \textbf{MCP}
    & \textbf{MRT}
    & \textbf{MG}$^\dagger$
    & \textbf{MINJA}$^\dagger$
    & \textbf{InjecAgent}$^\dagger$
    & \textbf{ASB}$^\dagger$
    & \textbf{AgentHarm} \\
    \midrule

    Sequential Monitor$^{a}$
    & $81.03^{\ddagger}$
    & $56.16^{\ddagger}$
    & --
    & $49.00^{\ddagger}$
    & 53.33
    & 50.00
    & 91.00
    & 69.23 \\

    Agent-Sentry L1$^{b}$
    & 84.44
    & 53.12
    & $51.90 \pm 30.99$
    & 91.38
    & 41.68
    & -- & -- & -- \\

    NeuroTaint$^{c}$
    & 72.53
    & 50.00
    & 50.00
    & 79.20
    & 49.24
    & 54.84
    & 50.00
    & 50.00 \\

    \midrule
    \textbf{GRAD}
    & \textbf{$95.16 \pm 0.63$}
    & \textbf{$62.10 \pm 1.76$}
    & \textbf{$74.00 \pm 8.59$}
    & \textbf{$99.82 \pm 0.31$}
    & \textbf{$94.91 \pm 0.93$}
    & --
    & --
    & -- \\
    \bottomrule
\end{tabular}%
}

\vspace{1mm}
\begin{minipage}{\linewidth}
    \footnotesize
    $^{a}$ DeepSeek native cumulative monitoring on Decomp;
    cumulative node-text and node-type adaptation
    elsewhere. The adapted scores come from the September 17 cumulative-prefix
    batch. MCP uses annotated node harm steps; MG uses the first harmful
    prefix. MINJA, InjecAgent, ASB, and AgentHarm use the maximum over
    all prefixes because harm deadlines are absent. The MRT adaptation
    was not executed in that batch.
    $^{b}$ Locally adapted L1 component, not the
    complete Agent-Sentry defense.
    $^{c}$ Propagation component without the
    counterfactual stage.

    \smallskip
    \noindent
    $^\ddagger$ Monitor scores for harmful trajectories
    use prefixes up to the annotated harm deadline,
    whereas benign trajectories use all prefixes.
    For MG, harmful trajectories are restricted to
    the first event. These results are not directly
    comparable to GRAD's full-trajectory detection.

    \smallskip
    \noindent
    The legacy unified GRAD reports mean$\pm$standard deviation
    across seeds 211--213. Version details appear in
    Appendix~\ref{app:detection_versions}.
    Agent-Sentry's MRT variation is across folds.
    $^\dagger$ MG/MINJA retain template/category
    confounds; derived InjecAgent/ASB contain strong
    shortcut cues. These columns do not establish
    attack generalization.
    AgentHarm concerns harmful-task behavior rather
    than indirect prompt injection.
\end{minipage}

\vspace{-2mm}
\end{wraptable}
\WFclear


\par
\begin{wraptable}{r}{\textwidth}
\centering
\caption{Historical extended binary-checker comparison.
Binary security-checking results for the evaluated
implementations.
Each entry reports TPR/FPR in percentages
($\uparrow$/$\downarrow$) under the specified
decision policy.
AP and AN denote all-positive and all-negative
predictions, respectively.
``--'' denotes an unavailable result.
Diagnostic datasets are marked with $^\dagger$
as in Table~\ref{tab:legacy_detection}.
}
\label{tab:legacy_binary}

\setlength{\tabcolsep}{5pt}
\renewcommand{\arraystretch}{1.12}

\resizebox{\linewidth}{!}{%
\begin{tabular}{lcccccccc}
    \toprule
    \multirow{2}{*}{\textbf{Method}}
    & \multicolumn{3}{c}{\textbf{Primary Detection Tasks}}
    & \multicolumn{2}{c}{\textbf{Memory Diagnostics}}
    & \multicolumn{3}{c}{\textbf{Additional Derived Tasks}} \\
    \cmidrule(lr){2-4}
    \cmidrule(lr){5-6}
    \cmidrule(lr){7-9}

    & \textbf{Decomp}
    & \textbf{MCP}
    & \textbf{MRT}
    & \textbf{MG}$^\dagger$
    & \textbf{MINJA}$^\dagger$
    & \textbf{InjecAgent}$^\dagger$
    & \textbf{ASB}$^\dagger$
    & \textbf{AgentHarm} \\
    \midrule

    AgentTrust + RiskChain$^{a}$
    & $0.0/0.0^{\mathrm{AN}}$
    & 3.5/0.0
    & 20.0/12.0
    & 30.0/0.0
    & $0.0/0.0^{\mathrm{AN}}$
    & $0.0/0.0^{\mathrm{AN}}$
    & $0.0/0.0^{\mathrm{AN}}$
    & $0.0/0.0^{\mathrm{AN}}$ \\

    AgentArmor$^{b}$
    & 98.9/91.4
    & 64.9/69.5
    & $100.0/100.0^{\mathrm{AP}}$
    & $100.0/100.0^{\mathrm{AP}}$
    & 33.3/47.0
    & $0.0/0.0^{\mathrm{AN}}$
    & $0.0/0.0^{\mathrm{AN}}$
    & $0.0/0.0^{\mathrm{AN}}$ \\

    AEGIS$^{c}$
    & 23.0/6.9
    & 54.4/54.2
    & $100.0/100.0^{\mathrm{AP}}$
    & 93.0/99.3
    & 0.0/6.1
    & 69.9/66.7
    & 96.0/100.0
    & 84.6/84.6 \\

    \midrule
    \textbf{GRAD}
    & 90.8/12.1
    & 33.4/13.3
    & 79.3/65.3
    & 68.4/0.2
    & 79.8/2.3
    & --
    & --
    & -- \\
    \bottomrule
\end{tabular}%
}

\vspace{1mm}
\begin{minipage}{\linewidth}
    \footnotesize
    $^{a}$ Rules-only configuration with
    chain-inclusive predictions and limited
    observable input fields.
    $^{b}$ Mechanism adaptation with an assumed
    registry and reconstructed dependencies.
    $^{c}$ Official classifier with a published
    policy schema and text-fallback input adaptation.

    \smallskip
    \noindent
    GRAD is included as a thresholded reference,
    using historical development-selected thresholds from the original
    unified batch, distinct from the paired G3-192 counts above.
    Test-set FPR is not constrained to equal the
    development target; operating points are
    not matched across methods.
    AP/AN indicate degenerate predictions in the
    evaluated configuration, not the effectiveness
    of the original method in general.
\end{minipage}

\vspace{-2mm}
\end{wraptable}
\WFclear

\par
\begin{wraptable}{r}{\textwidth}
\centering\small
\caption{Historical AuthGraph AgentDojo v1.1 protocol, DeepSeek Flash. Counts retain the original 18-run matched panel; not the P1/R1 or T3 cases. Runtime IRG construction is deterministic.}
\label{tab:authgraph_online}
\begin{adjustbox}{max width=\linewidth}
\begin{tabular}{lrrrrrr}
\toprule
Defense & Tool ASR & Text ASR & BU & UA & Errors & Degraded \\
\midrule
AuthGraph & 0/8 & 2/2 & 8/8 & 5/10 & 0/18 & 2/18 \\
\bottomrule
\end{tabular}
\end{adjustbox}
\end{wraptable}
\WFclear

\section{AgentDojo Attack Validity and Coverage}
\label{app:agentdojo}
These experiments diagnose attacks against the undefended agent
(\emph{None}); neither panel evaluates GRAD. Under the tested
DeepSeek Flash configuration, these panels measure undefended
attack success and payload delivery by scenario.

\subsection{Frozen Confirmation Set (P1)}
The frozen P1 manifest contains 35 independent user tasks, 70 attack
runs, and 35 benign runs. The agent uses DeepSeek Flash with thinking
disabled, temperature zero, and no retries. All 105 runs completed
without runtime errors. Attack success rate (ASR), benign task
completion (BU), and legitimate task completion under attack (UA)
are reported separately in Table~\ref{tab:app_agentdojo_p1}.

\par
\begin{wraptable}{r}{\textwidth}
\centering
\small
\caption{P1 undefended attack diagnostics. Rates are percentages.
Delivery denotes verified payload delivery. Multiple attack configurations
within a user task are not independent tasks.}
\label{tab:app_agentdojo_p1}
\begin{adjustbox}{max width=\linewidth}
\begin{tabular}{lrrrrrr}
\toprule
Suite & Attack runs & Benign runs & ASR & BU & UA & Delivery \\
\midrule
banking & 10 & 5 & 0.00 & 100.00 & 100.00 & 100.00 \\
slack & 20 & 10 & 5.00 & 100.00 & 85.00 & 100.00 \\
travel & 20 & 10 & 0.00 & 90.00 & 90.00 & 100.00 \\
workspace & 20 & 10 & 0.00 & 100.00 & 100.00 & 100.00 \\
\midrule
Overall & 70 & 35 & 1.43 & 97.14 & 92.86 & 100.00 \\
\bottomrule
\end{tabular}
\end{adjustbox}
\end{wraptable}
\WFclear

\subsection{Attack Development, Transfer, and Evaluator Audits}
The direct attack succeeded in 0/68 confirmation runs, while
\texttt{poison\_moved} succeeded in 1/2. No unexposed applicable task
remained for \texttt{poison\_forward}, so its transfer was untested.
Numeric-order configuration selection limited injection-goal coverage.
The banking \texttt{injection\_task\_4} evaluator was excluded uniformly
before evaluation after a deterministic reproduction identified a
conflict between its recurring-transaction requirement and the target
transaction state. Verified delivery establishes that a payload reached
the context, not that it constituted an effective attack.
The supplementary P1 archive at
\path{code/project/LOGS/final_table2/confirmation_batch1_20260919/}
contains \texttt{manifest\_p1.json},
\texttt{results\_stageC1.json}, and the exposure audit. Its single success
is Slack user task 16, injection task 3, with \texttt{poison\_moved}.
The aggregate reports run-level intervals; task-clustered uncertainty
and a complete success/near-miss comparison remain to be assembled.

\subsection{Official-Attack Adaptation Diagnostic (R1)}
R1 adapts the official \texttt{important\_instructions} payload to
DeepSeek Flash on benchmark version 1.2, selecting the first injection
task per suite that passed the evaluator sanity check and covering its
user tasks. The evaluated configuration uses this adapted payload
and the selected injection location in each suite.
Table~\ref{tab:app_agentdojo_r1} reports the recorded-result aggregation.

\par
\begin{wraptable}{r}{\textwidth}
\centering
\small
\caption{R1 undefended adaptation diagnostic. Rates are percentages.
The 94 recorded results include one budget-cap interruption; three
planned runs were not executed. ASR and UA retain the recorded-result
denominators. BU was not measured. Delivery check denotes automatic
string-based checking.}
\label{tab:app_agentdojo_r1}
\begin{adjustbox}{max width=\linewidth}
\begin{tabular}{lrrrrrr}
\toprule
Suite & Planned & Recorded & ASR & UA & Delivery check & Errors \\
\midrule
banking & 16 & 16 & 0.00 & 62.50 & 93.75 & 0 \\
slack & 21 & 21 & 0.00 & 76.19 & 100.00 & 0 \\
travel & 20 & 20 & 0.00 & 80.00 & 100.00 & 0 \\
workspace & 40 & 37 & 0.00 & 97.30 & 83.78 & 1 \\
\midrule
Overall & 97 & 94 & 0.00 & 82.98 & 92.55 & 1 \\
\bottomrule
\end{tabular}
\end{adjustbox}
\end{wraptable}
\WFclear

No attack success was observed among the 94 recorded results.
Automatic delivery checks passed for 87/94 records. P1 and R1
are separate undefended protocols; the defense comparisons use
the matched batches described in Appendix~\ref{app:runtime}.
Of 97 planned runs, 93 completed normally, one recorded run hit the
budget cap, and three were not executed. Treating all four unresolved
outcomes as either unsuccessful or successful gives a descriptive
full-plan ASR range of 0--4/97 (0--4.12\%), not a confidence interval.
The seven failed automatic delivery checks still require manual review.
No benign runs were collected for R1.

\section{Extended Ablations and Model Development}
\label{app:ablations}
\subsection{Full Ablation Matrix}
The final 512-character encoder comparisons appear in
Table~\ref{tab:app_grad512}; historical feature/relation ablations appear
in Table~\ref{tab:historical_features}. Matched G3 component removals are
reported separately in Table~\ref{tab:legacy_components}. The historical
feature ablations are not reruns on the final 512-character backbone.

\par
\begin{wraptable}{r}{\textwidth}
\centering\small
\caption{Historical pre-G3 feature and relation ablations: AUROC on [0,1], source-reported mean and SD across three seeds. These runs precede the final 512-character replication; source SD conventions are retained.}
\label{tab:historical_features}
\begin{adjustbox}{max width=\linewidth}
\begin{tabular}{lccc}
\toprule
Variant & Decomp & MCP & MRT \\
\midrule
Text only & $0.862\pm0.027$ & $0.617\pm0.021$ & $0.811\pm0.024$ \\
Structure only & $0.874\pm0.010$ & $0.675\pm0.016$ & $0.875\pm0.040$ \\
Chronological edges & $0.946\pm0.008$ & $0.655\pm0.025$ & $0.740\pm0.023$ \\
Without relation types & $0.954\pm0.008$ & $0.626\pm0.033$ & $0.787\pm0.056$ \\
Without keyword features & $0.963\pm0.011$ & $0.626\pm0.020$ & $0.848\pm0.048$ \\
Full joint encoder & $0.958\pm0.014$ & $0.641\pm0.019$ & $0.740\pm0.070$ \\
Without type semantics & $0.958\pm0.008$ & $0.626\pm0.009$ & --- \\
\bottomrule
\end{tabular}
\end{adjustbox}
\end{wraptable}
\WFclear

\subsection{Historical G3 Component Protocol}
\label{app:g3_ablation_details}
Table~\ref{tab:legacy_components} uses the existing component-attribution runs.
Each variant has three Decomp runs and 15 runs each for MCP and MRT.
Within each seed, fold AUROC is weighted by $n_+n_-$. The main table
reports the mean and sample standard deviation over seeds 211--213;
these are recomputed from the archived per-seed aggregates.

The LayerNorm ablation replaces the four input normalization modules
with identities. The auxiliary-loss ablation removes both training-only
heads, and the modality-dropout ablation sets its probability to zero.
Other G3 settings retain the frozen configuration. Structure only uses
permissions, position, node-type features, and the rebuilt graph;
its topology can retain text-derived construction cues. The regularized
structure model adds normalization, an attribute auxiliary objective,
and training-time attribute dropout. Full feature mappings and masking
rules accompany the implementation in Appendix~\ref{app:ssum_implementation}.

The original-encoder row uses the paired batch on Decomp and MCP and
reuses archived full-model runs on MRT. The historical G3 component records use
legacy evaluation splits; the reused held-out group evaluation is a
separate analysis. Group-bootstrap estimates based on pooled scores
are not interchangeable with the fold-weighted estimates reported here.
The final 512-character batch follows the input-window protocol in
Appendix~\ref{app:window512}.

\par
\begin{wraptable}{r}{\textwidth}
\centering\scriptsize
\caption{Historical GRAD-192 per-seed AUROC for SSUM ablations. Each cell lists seeds
211/212/213 in that order. Values are rounded for display.}
\label{tab:app_g3_seeds}
\begin{adjustbox}{max width=\linewidth}
\begin{tabular}{lccc}
\toprule
Variant & Decomp & MCP & MRT \\
\midrule
\textbf{Full encoder} & 0.9556/0.9823/0.9611 & 0.6618/0.6438/0.6085 & 0.9020/0.8160/0.9010 \\
w/o LayerNorm & 0.9660/0.9603/0.9716 & 0.5531/0.6439/0.5786 & 0.7980/0.7250/0.8190 \\
w/o auxiliary losses & 0.9700/0.9798/0.9800 & 0.6673/0.6456/0.6289 & 0.7170/0.8400/0.8500 \\
w/o modality dropout & 0.9554/0.9412/0.9325 & 0.6965/0.6529/0.6150 & 0.8270/0.6930/0.7250 \\
Original encoder & 0.9416/0.9758/0.9551 & 0.6664/0.6339/0.6226 & 0.8390/0.6860/0.6950 \\
Structure only & 0.8804/0.8827/0.8596 & 0.6766/0.6545/0.6926 & 0.8200/0.8920/0.9140 \\
Regularized structure only & 0.8766/0.8678/0.8786 & 0.6688/0.6486/0.6698 & 0.9140/0.8380/0.9320 \\
\bottomrule
\end{tabular}
\end{adjustbox}
\end{wraptable}
\WFclear

\subsection{Threshold and Score Diagnostics}
At the final G3-512 development thresholds, pooled development FPR
is 8.71\%/4.12\%/0.42\% on Decomp/MCP/MRT; corresponding test FPR
is 13.79\%/14.12\%/58.00\%. These are development-selected operating
points applied unchanged to test scores. Per-run thresholds and
confusion counts are recorded in the operating-point archive.

\section{Extended Continual-Learning Evaluation}
\label{app:continual}
\subsection{Task Matrices and Retention Metrics}
The final G3 panel uses five seeds and two task orders. Final
per-task accuracy is reported in Table~\ref{tab:g3_task_accuracy}; full
routing endpoints appear in Table~\ref{tab:g3_routing}. The reported
endpoints are final accuracy and forgetting. Offline joint training
is an accuracy reference trained on the combined data.
\subsection{Routing, Capacity, and Fair Comparisons}
Table~\ref{tab:historical_routing} reports the completed historical
MCP random/all-expert/centroid/task-oracle comparison. Final G3 results
for centroid/all-expert/task-oracle routing appear in
Table~\ref{tab:g3_routing}; final G3 random, learned, top-2, and fallback
routing results are unavailable. Training, replay, frozen state,
latency and storage are specified in Appendix~\ref{app:caa_snapshot}.

\par
\begin{wraptable}{r}{\textwidth}
\centering\small
\caption{Historical MCP routing comparison, distinct from the final G3 panels. Source-reported means and SD on [0,1].}
\label{tab:historical_routing}
\begin{adjustbox}{max width=\linewidth}
\begin{tabular}{lrrr}
\toprule
Router & ACC & Forgetting & Hit rate \\
\midrule
Random & $0.618\pm0.059$ & $0.069\pm0.047$ & 0.333 \\
All experts & $0.644\pm0.053$ & $0.015\pm0.019$ & --- \\
Centroid & $0.643\pm0.070$ & $0.010\pm0.013$ & 0.879 \\
Task oracle & $0.640\pm0.062$ & $0.000\pm0.000$ & 1.000 \\
\bottomrule
\end{tabular}
\end{adjustbox}
\end{wraptable}
\WFclear

\section{Runtime and Cross-Backbone Diagnostics}
\label{app:runtime}
\subsection{Joint-Transfer Deployment}
\par
\begin{wraptable}{r}{\textwidth}
\centering\small
\caption{Detector transfer AUROC (\%). T3 initializes from the Decomp-trained
GRAD checkpoint and jointly trains on Decomp and closed-loop gate data.
The Table~\ref{tab:detection} initialization is a retrained fold-0, seed-211 checkpoint.}
\label{tab:gate_lineage}
\begin{adjustbox}{max width=\linewidth}
\begin{tabular}{lccc}
\toprule
Detector & Gate dev & Gate test & Decomp test \\
\midrule
From-scratch loop gate & 94.12 & 94.77 & 50.31 \\
Decomp-trained GRAD & 43.53 & 68.60 & 94.28 \\
GRAD joint transfer (T3) & \textbf{97.41} & \textbf{96.08} & \textbf{96.88} \\
\bottomrule
\end{tabular}
\end{adjustbox}
\end{wraptable}
\WFclear

The deployed T3 detector initializes from the retrained Table~\ref{tab:detection} Decomp
fold-0, seed-211 checkpoint, then trains on 1,397 Decomp and 172 gate
training examples for 100 epochs at learning rate $10^{-3}$.
Its development-selected threshold is 0.0226; checkpoint SHA-256 is
\texttt{e0f79c22...b1f1712}. Node and edge windows are 512 and 160
characters. The gate/trainer tensor-identity check passes for all six
fields, with maximum score difference $2.9\times10^{-5}$.

The declared deployment conditions use gate-development AUROC,
gate-test AUROC (both at least 0.90), and development fixed-FPR-10
F1 (at least 0.75); among qualifying configurations, selection uses
highest development AUROC. T3 is the only qualifying configuration,
with development F1 0.938. The gate-test split served as a pre-specified
deployment acceptance set rather than as an untouched final test set.
Gate-test performance did not rank qualifying configurations and acted
only as an acceptance criterion; nevertheless, because it affected
qualification, we do not treat the resulting gate-test AUROC as an
independent held-out estimate. The 96.08\% value is a descriptive
performance measure of the deployed detector. The transfer comparison
is reported as exploratory. The closed-loop evaluation reuses the earlier case panel and is
reported as exploratory. The Decomp value 96.88\% is a single
adapted-checkpoint measurement; Table~\ref{tab:detection}'s 97.19\% summarizes three
seeds.

The static detection and CAA experiments use their own development
selection protocols and checkpoints. The T3 batch reuses the earlier
closed-loop case panel and intervention protocol, deploying the
adapted detector as \texttt{g3x}. All 240 runs execute on the GPU server, with zero
infrastructure errors. The report and aggregate are archived under
\path{code/project/LOGS/final_table4_g3trans/}.

\par
\begin{wraptable}{r}{\textwidth}
\centering\small
\caption{Closed-loop results across the two fixed-case batches (\%).
Each arm has 20 attack and 20 benign runs per backbone per batch.
Earlier GRAD uses the from-scratch loop gate; current GRAD uses T3.
ASR, BU, and UA denote attack success, benign completion, and legitimate
completion under attack.}
\label{tab:runtime_batches}
\begin{adjustbox}{max width=\linewidth}
\begin{tabular}{llrrrrrr}
\toprule
 & & \multicolumn{3}{c}{Earlier batch} & \multicolumn{3}{c}{T3 batch} \\
Backbone & Defense & ASR & BU & UA & ASR & BU & UA \\
\midrule
DeepSeek & None & 35 & 100 & 45 & 35 & 100 & 50 \\
 & Monitor & 0 & 60 & 25 & 0 & 50 & 25 \\
 & GRAD & 0 & 85 & 65 & 0 & 75 & 80 \\
GLM & None & 35 & 100 & 55 & 30 & 90 & 45 \\
 & Monitor & 15 & 45 & 5 & 0 & 50 & 0 \\
 & GRAD & 20 & 95 & 45 & 10 & 40 & 30 \\
\bottomrule
\end{tabular}
\end{adjustbox}
\end{wraptable}
\WFclear

\par
\begin{wraptable}{r}{\textwidth}
\centering\small
\caption{Exact paired McNemar tests against the same-backbone, same-batch None
arm. Each pair lists discordant counts $(b,c)$ and the unadjusted
two-sided $p$-value; gate batches are reported separately.}
\label{tab:runtime_paired}
\begin{adjustbox}{max width=\linewidth}
\begin{tabular}{llrrrr}
\toprule
 & & \multicolumn{2}{c}{Harm} & \multicolumn{2}{c}{Benign completion} \\
Backbone & Defense & $(b,c)$ & $p$ & $(b,c)$ & $p$ \\
\midrule
\multicolumn{6}{l}{\textit{Current T3 batch}} \\
DeepSeek & Monitor & (7,0) & 0.0156 & (10,0) & 0.0020 \\
 & GRAD & (7,0) & 0.0156 & (5,0) & 0.0625 \\
GLM & Monitor & (6,0) & 0.0312 & (8,0) & 0.0078 \\
 & GRAD & (6,2) & 0.2891 & (10,0) & 0.0020 \\
\midrule
\multicolumn{6}{l}{\textit{Earlier from-scratch gate batch}} \\
DeepSeek & Monitor & (7,0) & 0.0156 & (8,0) & 0.0078 \\
 & GRAD & (7,0) & 0.0156 & (3,0) & 0.2500 \\
GLM & Monitor & (4,0) & 0.1250 & (11,0) & 0.0010 \\
 & GRAD & (4,1) & 0.3750 & (1,0) & 1.0000 \\
\bottomrule
\end{tabular}
\end{adjustbox}
\end{wraptable}
\WFclear

Across batches, the unchanged monitor's GLM ASR shifts from 15\%
to 0\%, while GRAD's GLM BU shifts from 95\% to 40\%.
The two batches combine stochastic agent execution with a changed gate.
Paired tests compare defenses within each batch.

\subsection{Earlier From-Scratch Gate Batch}
The earlier closed-loop results in Table~\ref{tab:runtime_batches} come from the
controlled two-session online environment, not AgentDojo or a prefix
replay. The agent models are DeepSeek Flash and GLM-5.3-Flash. The
Sequential Monitor uses the fixed DeepSeek Flash judge, its original
binary-intention prompt and a 0.5 decision threshold. The GRAD arm
uses the G3-512 \texttt{ModalityNormGNN} gate trained under the frozen
v2 gate protocol, with a 512-character node window, 160-character edge
window and a development-selected fixed-FPR/F1 threshold of 0.9999915.
The checkpoint SHA-256 begins \texttt{f79f477b} and ends
\texttt{bdd3d}; its full value is in the run manifest.
It is held fixed across agent models without target-specific
recalibration. A blocked action triggers at most two replanning attempts.

The frozen plan SHA-256 begins \texttt{c6a89f27} and ends
\texttt{71de}; its full value is in the run report.
It contains four attack families, benign and attack conditions, and
five repeats, giving 20 runs per condition and arm in each agent-model
panel. None, Sequential Monitor, and GRAD were rerun within the same
batch (240 runs total). Sanitizer was not rerun with this G3-512 gate
and is omitted from the matched comparison. This batch uses the fixed
case panel described in Appendix~\ref{app:runtime}.
\subsection{Errors, Interventions, and Label Revisions}
All 240 runs completed with zero infrastructure errors. The matched
GRAD arms recorded at least one block in 10/40 DeepSeek and 3/40 GLM
runs; Sequential Monitor did so in 31/40 and 22/40, respectively.
For DeepSeek, harmful outcomes were 7/20 for None and 0/20 for both
defenses. GRAD benign completion is 17/20,
versus 12/20 for Sequential Monitor and 20/20 for None. For GLM,
harmful outcomes were 7/20 for None, 3/20 for Sequential Monitor,
and 4/20 for GRAD. Paired tests appear in
Table~\ref{tab:runtime_paired}. Benign completion is 19/20 for GRAD, 9/20 for Sequential Monitor, and 20/20
for None. Under attack, legitimate completion is 13/20, 5/20, and
9/20 for DeepSeek GRAD, Monitor, and None, respectively, and 9/20,
1/20, and 11/20 in the same order for GLM. These denominators preserve
errors as failures; no error occurred in this batch. The GLM GRAD
harmful outcomes are concentrated in the shopping family (4/5).
In the GLM benign-condition GRAD arm, one workspace-family run
(\texttt{zeta}, repeat 2) has \texttt{independent\_harm=true} and
\texttt{independent\_completion=false} with no block or runtime error.
Its gate score is 0.997315, below the frozen threshold 0.999991536.
The corresponding None and Monitor runs complete the legitimate task.
This historical case was flagged for evaluator and trace review.

The matched-batch source is
\path{code/project/LOGS/final_table4_g3/main/results.json}; the predeclared plan,
aggregate, cell provenance, and report are stored under
\path{code/project/LOGS/final_table4_g3/}. The gate-vs-trainer input test found
six graph-tensor fields exactly equal; on 42 development rows, score
parity had maximum absolute difference $2.3\times10^{-4}$ between
batch-one gate inference and the training batch. The first 38 runs
were executed locally and the remaining runs on the GPU server after verifying identical code, prompt,
and gate hashes. The recorded timers consequently combine two hosts
and should not be interpreted as a single-hardware benchmark.
These host and timing details apply to the earlier batch only;
Table~\ref{tab:online_cost} uses the server-only T3 batch.

\section{Efficiency and Reproducibility}
\label{app:reproducibility}
Experiments were conducted on a server equipped with eight NVIDIA
GeForce RTX 3090 GPUs. The T3 closed-loop batch ran entirely on the server; the earlier
from-scratch batch used the local host for its first 38 runs.

The G3-512 gate checkpoint occupies 2.6\,MB and performs in-process
inference. The continual-learning snapshot pool occupies 7.74\,MB;
its training and routing costs are reported in
Appendix~\ref{app:caa_snapshot}. The closed-loop batch completed
240/240 runs with no infrastructure errors. Its logged gate-forward
and monitor-call latencies appear in Table~\ref{tab:online_cost};
the cost of local compute is not converted into a currency estimate.
At the batch's fixed conversion of 7.15 CNY per US dollar, the API
ledgers correspond to approximately US\$0.0872 for 922 DeepSeek calls
and US\$0.0292 for 836 GLM calls, or US\$0.1164 across the T3 batch
(CNY 0.8322 in the original ledgers).
The underlying billed amounts remain in the source ledgers. These
costs include agent generation, judge calls where applicable, and
intervention-induced replanning.
The GRAD timer starts after graph construction and tensor preparation
and covers only model forward inference and probability conversion.
The monitor timer covers the hosted judgment call. Agent generation
and replanning are outside both timers. Detailed full-path decision
timings, per-stage trajectory timings, peak host memory,
and monitoring-token counts were not recorded in the verified
aggregate and are not inferred here.

The machine-readable closed-loop results, plan, cell provenance,
checkpoint identity, run metadata, and aggregation script are archived
in the supplementary code package under \path{code/project/LOGS/final_table4_g3trans/} and
\path{code/project/code/repro/final_table4/}. The plan and checkpoint hashes are
listed in Appendix~\ref{app:runtime}; the gate seed is 211.

\section{Additional Method Details}
\label{app:method_details}

\subsection{Attack Formulation}
\label{app:threat_details}

The attacker seeks to induce an agent $\pi$ to achieve
a harmful objective $g$ that violates the application
security policy. We consider decomposition attacks
and memory poisoning, which challenge local safeguards
by distributing attack evidence across subtasks or
separating malicious information insertion from its
subsequent use.

\noindent\textbf{Decomposition attacks.}
Following \citet{yueh2026monitoring}, the attacker
constructs a sequence of seemingly benign subrequests
$Q_g=(q_1,\ldots,q_L)$ whose combined execution
serves $g$. Each subrequest changes the
agent--environment state:
\begin{equation}
    \sigma_{\ell}
    =
    \mathcal{F}_{\pi}(\sigma_{\ell-1},q_{\ell}),
    \qquad \ell=1,\ldots,L.
\end{equation}
Earlier subrequests may produce information or
artifacts used by later ones. The attacker controls
the wording and order of these subrequests, while
each subrequest may induce multiple monitored events.

\noindent\textbf{Memory poisoning attacks.}
The attacker contaminates a memory or knowledge
store $\mathcal{M}$ through direct insertion or
interactions that influence memory updates:
\begin{equation}
    \mathcal{M}^{p}
    =
    \begin{cases}
        \mathcal{M}\cup\mathcal{P},
        & \text{direct insertion},\\
        \mathcal{W}_{\pi}(\mathcal{M},\mathcal{I}_a),
        & \text{interaction-induced storage}.
    \end{cases}
\end{equation}
Here, $\mathcal{P}$ denotes injected records and
$\mathcal{I}_a$ denotes attacker-controlled
interactions. These settings correspond to partial
write access as in AgentPoison
\citep{chen2024agentpoison}, and query-induced
memory contamination as in MINJA
\citep{dong2025memory}, respectively.
A subsequent victim request $q^v$ retrieves
$Z^p=\operatorname{Retrieve}(q^v,\mathcal{M}^p)$,
allowing poisoned information to influence planning
and tool execution.

For either attack, the attacker chooses an admissible
intervention $\delta\in\Delta(b)$ under budget $b$ to
maximize harmful-goal completion:
\begin{equation}
    \max_{\delta\in\Delta(b)}
    \mathbb{E}\!\left[
        \mathcal{V}_g
        \bigl(\sigma_T^{\pi,\mathrm{GRAD}}(\delta)\bigr)
    \right],
\end{equation}
where $\mathcal{V}_g$ indicates whether $g$ is
achieved in the defended execution. The monitoring
infrastructure, detector parameters, and labeled
adaptation data are trusted. Attack access and
budgets are specified for each experimental setting.

\subsection{Interpretation and Evaluation Scope}
\label{app:interpretation_scope}
SSUM aggregates observed operations along candidate relations
constructed by the rules in Appendix~\ref{app:ssum_implementation}.
CAA adapts an external detector through frozen expert snapshots;
system predictions combine each expert's function with the router's
selection as the collection grows.

Offline graph labels supervise the evaluated graph rather than every
constituent operation. Task-stream evaluation uses the expert collection,
whereas closed-loop intervention uses a separately trained standalone
detector. These experiments assess distinct system configurations.

\subsection{SSUM Implementation and Training}
\label{app:ssum_implementation}
\noindent\textbf{Graph construction.}
Events are stably sorted by their recorded turn field. Adjacent nodes
receive forward sequence edges and eligible type/lexical relations.
Additional candidate edges span sorted node-index differences of two
to five. These include memory-write/read pairs and eligible data-flow
and sink-trigger pairs. Edges are deduplicated by source, target, and
relation type, allowing multiple relations per node pair. Heuristic
weights are constants assigned by the construction rules. The G3
configuration uses rebuilt edges with copied edge descriptions.

\noindent\textbf{Architecture.}
GRAD uses a 512-character node window and a 160-character edge
window. Node and edge character CNNs each output 64 dimensions. Node base
features contain 12 permission entries, one normalized position, and
nine lexical indicators. Edge base features contain a relation one-hot
vector and one heuristic weight. Four independent LayerNorm modules
normalize node base, node text, edge base, and edge text blocks before
linear projection to dimension 192. The encoder uses one GINE layer
and two four-head TransformerConv layers. Each layer applies BatchNorm,
GELU, dropout at rate 0.35, and a residual connection. Attention pooling
clamps its denominator at $10^{-8}$. The joint classification head has
a 96-dimensional hidden layer, GELU, dropout, and a scalar output.

\noindent\textbf{Modality masks and auxiliary heads.}
\label{app:ssum_masks}
Two independent Bernoulli masks are drawn per graph, each with drop
probability 0.25, and shared across its nodes. One zeros permissions
and position; the other zeros lexical indicators and node character
embeddings. Masking precedes LayerNorm. Edge inputs and topology remain
active. The text auxiliary branch projects the masked, normalized
64-dimensional node character embedding to dimension 192 with GELU.
The attribute branch projects the masked 13-dimensional permission
and position vector to dimension 192 with GELU. Each branch has its
own sigmoid-attention pool and linear classification head. Auxiliary
branches execute during training only.

\noindent\textbf{Loss weighting.}
\label{app:ssum_loss}
Writing $q$ for the protocol's positive-class weight, the per-example
loss is $\ell_q(y,p)=-q y\log p-(1-y)\log(1-p)$. Equation~\ref{eq:detection_objective}
uses $w(G,y)=s(G)[q y+(1-y)]$, where $s(G)$ is the sample weight.
Auxiliary losses average $\ell_q$ with unit sample weights. Each
auxiliary term has coefficient 0.1. Training samples one mask pair
per graph to estimate Eq.~\ref{eq:ssum_training}.

\subsection{Training and Prediction Procedure}
\label{app:grad_algorithm}
Algorithm~\ref{alg:grad} gives the full expert-training, prediction,
and standalone runtime-gating procedure. The runtime gate uses a
separately trained SSUM detector rather than the expert collection.

\begin{algorithm}[t]
\small
\caption{GRAD: Training and Prediction}
\label{alg:grad}
\DontPrintSemicolon
\KwIn{Labeled task stream $\mathcal{D}_1,\ldots,\mathcal{D}_R$;
 graph constructor $\Phi$; development data.}
\KwOut{Expert collection
 $\mathcal{A}_R=\{(\theta_j^\star,\phi_j,\boldsymbol{\mu}_j)\}_{j=1}^{R}$.}
Initialize the working backbone, empty replay buffer and expert collection\;
\For{$r=1,\ldots,R$}{
 Update the working backbone with current data and historical replay
 (Appendix~\ref{app:caa_snapshot})\;
 Copy the selected backbone to frozen snapshot $\theta_r^\star$\;
 Train adapter $A_r$ and head $H_r$ on $\mathcal{D}_r$
 (Eq.~\ref{eq:expert_learning})\;
 Store $(\theta_r^\star,\phi_r,\boldsymbol{\mu}_r)$ in $\mathcal{A}_r$\;
 Update the replay buffer with current examples and cached expert probabilities\;
}
\textbf{Predict with the expert collection}\;
For input $G$, select $j^\star(G)$ and return $p_{j^\star(G)}(G)$
 using Eq.~\ref{eq:antibody_routing}\;
\textbf{Apply the standalone runtime detector}\;
Given observed history $\mathcal{E}_{<t}$ and proposed action $a_t$,
 construct $G_t=\Phi(\mathcal{E}_{<t},a_t)$\;
Score $G_t$ with the runtime SSUM detector; intervene when its score
 reaches the development-selected threshold $\tau$\;
\end{algorithm}

\subsection{CAA Implementation}
\label{app:caa_implementation}
Each frozen encoder snapshot includes its normalization and pooling
parameters; BatchNorm running statistics are fixed during expert training. An adapter
is inserted at the output of the final TransformerConv, before the
BatchNorm, activation, dropout, and residual update of
Eq.~\ref{eq:message_passing}. Its bias-free down/up projections have
bottleneck dimension $b=8$, with the up-projection initialized to zero.
Each added expert uses a linear classification head. The snapshot-specific pooling
operator applied to adapted node states produces $\mathbf{g}_r(G)$.

\noindent\textbf{Routing signature.}
\label{app:caa_routing}
The signature uses a normalized node-type histogram and scaled graph counts:
\begin{equation*}
\mathbf{s}(G)=[\operatorname{Hist}_{\mathrm{type}}(G)
 \Vert |V|/50\Vert |E_{\mathrm{input}}|/50],\qquad
\boldsymbol{\mu}_r=\frac{1}{|\mathcal{D}_r|}
 \sum_{(G,y)\in\mathcal{D}_r}\mathbf{s}(G).
\end{equation*}
Here $E_{\mathrm{input}}$ is the supplied edge set before SSUM edge
reconstruction. Unknown node types map to the other category.
Euclidean distance selects a centroid; ties follow stored task order.
Task identifiers group training examples, while inference uses the
observable signature. Routing compares all stored centroids and
executes the selected expert. Expert storage grows with the task count.

\noindent\textbf{Version mapping.}
The final reported snapshot procedure is specified in
Appendix~\ref{app:caa_snapshot}. The stage-0 baseline shares one frozen encoder across experts;
snapshot CAA uses a separate frozen encoder copy per expert, including
at the first stage.

\section{Historical Continual-Learning Reference}
\label{app:historical_cl}
These results retain the earlier experimental configuration and are kept
separate from the final GRAD CAA evaluation.


\par
\begin{wraptable}{r}{\textwidth}
\centering
    \caption{
    Historical continual-learning batch before the final CAA evaluation, reported as mean $\pm$
    standard deviation.
    MCP in-domain evaluation uses 40 runs: 20 seeds $\times$ two task orders;
    cross-dataset evaluation uses 10 runs: five seeds $\times$ two task orders.
    The cross-dataset stream contains MCP, MRT, Decomp,
    and MG-derived tasks.
    Forgetting follows
    $F_T=\frac{1}{T}\sum_{j=1}^{T}
    [\max_{t=j,\ldots,T}a_{t,j}-a_{T,j}]$.
    Bold marks the best observed mean among the listed
    continual-learning methods.
    Joint training is a separate finite-budget offline reference; its
    forgetting cells are omitted because they are not sequential-retention results.
    }
    \label{tab:continual_legacy}

    \setlength{\tabcolsep}{8pt}
    \renewcommand{\arraystretch}{1.15}

    \resizebox{\linewidth}{!}{%
    \begin{tabular}{lcccc}
        \toprule
        \multirow{2}{*}{\textbf{Method}}
        & \multicolumn{2}{c}{\textbf{MCP In-domain}}
        & \multicolumn{2}{c}{\textbf{Cross-dataset}} \\
        \cmidrule(lr){2-3}
        \cmidrule(lr){4-5}

        & ACC$\uparrow$
        & $F_T\downarrow$
        & ACC$\uparrow$
        & $F_T\downarrow$ \\
        \midrule

        \rowcolor{gray!15}
        Sequential fine-tuning
        & $0.568\pm0.106$
        & $0.110\pm0.106$
        & $0.684\pm0.087$
        & $0.095\pm0.054$ \\

        Replay
        & $0.568\pm0.122$
        & $0.079\pm0.085$
        & $\mathbf{0.760\pm0.053}$
        & $0.044\pm0.034$ \\

        EWC
        & $0.574\pm0.092$
        & $0.093\pm0.079$
        & $0.630\pm0.070$
        & $0.162\pm0.068$ \\

        \midrule
        \textbf{GRAD antibody experts}
        & $\mathbf{0.643\pm0.070}$
        & $\mathbf{0.010\pm0.013}$
        & $0.720\pm0.035$
        & $\mathbf{0.000\pm0.001}$ \\

        \midrule
        Joint training
        & $0.543\pm0.114$
        & ---
        & $0.727\pm0.092$
        & --- \\
        \bottomrule
    \end{tabular}%
    }
    \vspace{-2mm}
\end{wraptable}
\WFclear

\section{Limitations and Evaluation Scope}
\label{app:limitations}
The current datasets contain finite execution histories. Longer agent
sessions increase graph size and the amount of evidence that must be
retained. Incremental graph processing and selective history retention
are directions for evaluating this scaling behavior. Integration into
another agent framework also requires mapping its tool and memory events
to the input schema.

The final 512-character encoder comparison appears in
Appendix~\ref{app:window512}; historical 192-character results are
identified by version in their table captions. Three-seed summaries
report run variability, with task- or family-grouped paired analyses.

Continual evaluation depends on the task stream, training allocation,
expert initialization, and routing policy. Expanding the library also
increases storage and selection costs. The final panels use G3-512; the earlier backbone is reported separately
in Appendix~\ref{app:historical_cl}.

Runtime evaluation depends on the attack signal, agent configuration,
and intervention behavior. The AgentDojo diagnostics document sparse
undefended attack success in the tested settings. Frozen-detector
transfer and recalibration answer different deployment questions, and
API errors or interrupted runs require explicit denominator accounting.

\section{Input-Window Replication}
\label{app:window512}
Table~\ref{tab:app_grad512} collects the current four-arm results.

\par
\begin{wraptable}{r}{\textwidth}
\centering\small
\caption{Frozen-scheme replication on legacy test splits: AUROC (\%),
mean and sample SD across seeds 211--213. All four arms were retrained
in the same batch. GRAD and the text-only arm use 512-character node
windows; the GRAD-192 arm uses 192 characters.}
\label{tab:app_grad512}
\begin{adjustbox}{max width=\linewidth}
\begin{tabular}{lccc}
\toprule
Arm & Decomp & MCP & MRT \\
\midrule
\textbf{GRAD} & $\mathbf{97.19\pm0.99}$ & $77.59\pm2.01$ & $86.60\pm5.13$ \\
Graph only & $86.37\pm0.79$ & $67.41\pm0.89$ & $86.07\pm3.95$ \\
Text only & $86.83\pm0.48$ & $\mathbf{79.89\pm2.72}$ & $82.87\pm6.40$ \\
GRAD-192 & $96.40\pm0.53$ & $64.56\pm3.98$ & $\mathbf{86.83\pm3.52}$ \\
\bottomrule
\end{tabular}
\end{adjustbox}
\end{wraptable}
\WFclear

The EXP-2026W39-004 batch contains 132 completed runs: four arms,
11 dataset/fold cells, and seeds 211--213. Two runs failed with GPU
out-of-memory errors and were rerun with the same configuration.
The replication follows the split-reuse timeline in
Appendix~\ref{app:data}.

GRAD uses A\_moddrop, auxiliary-loss coefficient 0.1, modality-drop
probability 0.25, and a 512-character node window. The edge window
remains 160 characters. Training uses 100 epochs, patience 25, batch
size 48, learning rate $10^{-3}$, and the existing development-selection
protocol. The fresh 192-character reference retains the same G3 design.
TXT-512 uses the text-only implementation with a matched node window;
STRUCT is the fresh structure-only run. The main tables' sample SDs
are calculated from the report's per-seed, fold-weighted AUROCs.

On Decomp and MCP, paired bootstrap intervals for GRAD minus the
graph-only encoder are positive in all three seeds. On MCP, differences
from the text-only encoder have intervals spanning zero. MRT is assessed with task-level sign tests; neither
single-channel comparison reaches significance. Within-task MRT bootstrap
intervals are degenerate because each task cell contains one group.

The preregistration compared GRAD with both single-channel encoders
on all three datasets. The observed dataset-specific results led to
branch (b): adopt the 512-character input repair as the backbone for
subsequent antibody experiments. The GRAD-192 versus GRAD comparison controls encoder
design and training configuration within this batch; legacy-text
comparisons also change the run batch.

\subsection{Historical Decomp v2 Evaluation}
The earlier 192-character branch was evaluated on 665 newly converted
Decomp graphs, with 334 base-id groups. Its three-seed AUROCs are
reported in Table~\ref{tab:decomp_v2_history}; the dispersion follows
the archived report. G3 minus structure-only paired intervals include
zero in every seed. This batch is distinct from the final 512-character
legacy-split replication.
\par
\begin{wraptable}{r}{\textwidth}
\centering\small
\caption{Historical Decomp v2 evaluation: AUROC and PR-AUC on [0,1].
AUROC dispersion is the source-reported three-seed SD.}
\label{tab:decomp_v2_history}
\begin{adjustbox}{max width=\linewidth}
\begin{tabular}{lrr}
\toprule
Method & AUROC & PR-AUC \\
\midrule
Original encoder & $0.7886\pm0.0286$ & 0.8979 \\
G3\_p25 & $0.8513\pm0.0215$ & 0.9361 \\
Structure only & $0.8525\pm0.0030$ & 0.9448 \\
Regularized structure & $0.8501\pm0.0077$ & 0.9446 \\
\bottomrule
\end{tabular}
\end{adjustbox}
\end{wraptable}
\WFclear

\subsection{Historical Training-Component Ablations}
\par
\begin{wraptable}{r}{\textwidth}
\centering\small
\caption{Historical GRAD-192 component ablations: AUROC (\%), mean and sample standard
 deviation across seeds 211--213. Component removals use G3\_p25 as
 reference. Protocol and variant definitions appear in
 Appendix~\ref{app:g3_ablation_details}.}
\label{tab:legacy_components}
\begin{adjustbox}{max width=\linewidth}
\begin{tabular}{lccc}
\toprule
Variant & Decomp & MCP & MRT \\
\midrule
\textbf{Full encoder} & $96.63\pm1.41$ & $63.80\pm2.71$ & $87.30\pm4.94$ \\
w/o LayerNorm & $96.59\pm0.56$ & $59.19\pm4.69$ & $78.07\pm4.93$ \\
w/o auxiliary losses & $\mathbf{97.66\pm0.57}$ & $64.73\pm1.93$ & $80.23\pm7.41$ \\
w/o modality dropout & $94.31\pm1.16$ & $65.48\pm4.08$ & $74.83\pm7.00$ \\
\midrule
Original encoder & $95.75\pm1.72$ & $64.10\pm2.28$ & $74.00\pm8.59$ \\
Graph only & $87.43\pm1.27$ & $\mathbf{67.46\pm1.91}$ & $87.53\pm4.92$ \\
Regularized graph only & $87.43\pm0.58$ & $66.24\pm1.20$ & $\mathbf{89.47\pm4.99}$ \\
\bottomrule
\end{tabular}
\end{adjustbox}
\end{wraptable}
\WFclear

\subsection{Current-Version Operating Points and Artifact Coverage}
\label{app:grad512_operating_points}
Each run applies its development-selected threshold to stored test
scores using $\hat y=\mathbb{1}[s\geq\tau]$. The aggregation covers
3 Decomp runs, 15 MCP runs, and 15 MRT runs across seeds 211--213.
Table~\ref{tab:grad512_counts} reports pooled test counts.
The corresponding pooled development FPRs are 8.71\%, 4.12\%, and
0.42\%; MRT exhibits a large development-to-test calibration shift.
These confirmation-batch operating points are reconstructed from
archived per-example scores. Model checkpoints were not saved in
those original confirmation-run directories.

\par
\begin{wraptable}{r}{\textwidth}
\centering\small
\caption{GRAD pooled test confusion counts at per-run development
thresholds. Repeated seeds contribute repeated observations.}
\label{tab:grad512_counts}
\begin{adjustbox}{max width=\linewidth}
\begin{tabular}{lrrrrrr}
\toprule
Dataset & TP & FP & TN & FN & TPR (\%) & FPR (\%) \\
\midrule
Decomp & 498 & 24 & 150 & 24 & 95.4 & 13.8 \\
MCP & 178 & 50 & 304 & 164 & 52.0 & 14.1 \\
MRT & 125 & 87 & 63 & 25 & 83.3 & 58.0 \\
\bottomrule
\end{tabular}
\end{adjustbox}
\end{wraptable}
\WFclear

\subsection{Online Cost Measurement Protocol}
\label{app:online_cost_protocol}
Table~\ref{tab:online_cost} reports timing data from the matched
240-run batch. For GRAD, the logged \texttt{latency\_ms} is
measured inside \texttt{G3Gate.score\_nodes}, after graph construction
and tensor preparation; it covers the model forward pass and sigmoid.
For Sequential Monitor, the timer surrounds the hosted judge call.
The table reports these two scoring operations. Agent generation
and replanning lie outside both recorded timers.

The total API ledger includes base-agent generation, judge calls for
Sequential Monitor, and intervention-induced replanning. The local
gate adds no monitoring API request, but still uses local compute;
its compute cost is not monetized. The T3 batch ran entirely on
the server. The earlier batch combined local and server execution. Full-path latency, monitoring
token counts, and peak detector memory remain unmeasured in this batch.

\section{Snapshot-Based Antibody Evaluation}
\label{app:caa_snapshot}
The formal cross-dataset panel uses five seeds and two orders over
Decomp/MCP/MRT. A mutable backbone is trained at each stage with current
examples and a buffer of 100 historical examples with cached probabilities.
Replay uses a binary-cross-entropy distillation term of weight 1.0.
After development selection, the backbone is copied and frozen; a new
adapter and linear head are trained on that snapshot and current data.
The initial task follows the same sequence. Historical experts retain
their own snapshots, adapters and heads. The newly trained expert's
probabilities populate the replay buffer. The router stores the current
training-signature centroid. The stage-0 arm retains the original fixed
initial backbone and is a separate ablation.

Paired comparisons average task orders within seeds and bootstrap seed
clusters. On the G3 cross-dataset panel, snapshot minus Replay yields
ACC +0.1308 [0.0911,0.1910] and forgetting -0.0551 [-0.1185,-0.0169].
Snapshot minus stage-0 yields ACC +0.0949 [0.0834,0.1102] and forgetting
+0.0030 [-0.0024,0.0085].
On the family-isolated panel, snapshot minus Replay yields ACC +0.1191
[0.0395,0.1930] and forgetting -0.0127 [-0.0534,0.0315]. Stage-0 minus
Replay yields ACC +0.1398 [0.0954,0.2013] and forgetting -0.0412
[-0.0719,-0.0129]. Snapshot minus stage-0 yields ACC -0.0207
[-0.1552,0.0891] and forgetting +0.0285 [0.0034,0.0667].
The family-isolated snapshot forgetting advantage over Replay is not
statistically established. These comparisons use the G3 formal aggregates and the
same backbone implementation as the static G3 detector.

\par
\begin{wraptable}{r}{\textwidth}
\centering\small
\caption{Final G3-512 routing endpoints, five seeds and two orders per panel. Accuracy is percent; forgetting is on [0,1].}
\label{tab:g3_routing}
\begin{adjustbox}{max width=\linewidth}
\begin{tabular}{lrrrr}
\toprule
Method & DMM ACC & DMM $F_R$ & Family ACC & Family $F_R$ \\
\midrule
Sequential FT & 56.43 & 0.1251 & 45.62 & 0.0797 \\
Replay & 64.78 & 0.0629 & 39.21 & 0.0454 \\
Stage-0 centroid & 68.37 & 0.0048 & 53.18 & 0.0042 \\
Stage-0 all & 63.89 & 0.0047 & 44.22 & 0.0326 \\
Stage-0 task oracle & 68.31 & 0.0000 & 49.36 & 0.0000 \\
Snapshot centroid & 77.86 & 0.0079 & 51.11 & 0.0327 \\
Snapshot all & 71.58 & 0.0145 & 41.98 & 0.0562 \\
Snapshot task oracle & 79.86 & 0.0000 & 49.40 & 0.0000 \\
\bottomrule
\end{tabular}
\end{adjustbox}
\end{wraptable}
\WFclear

\par
\begin{wraptable}{r}{\textwidth}
\centering\small
\caption{Final G3 per-task accuracy (percent), averaged over five seeds and two orders. Rows correspond to the same panels as Table~\ref{tab:g3_routing}.}
\label{tab:g3_task_accuracy}
\begin{adjustbox}{max width=\linewidth}
\begin{tabular}{llrrr}
\toprule
Panel & Method & Task 1 & Task 2 & Task 3 \\
\midrule
Decomp/MCP/MRT & Sequential FT & 61.67 & 46.00 & 61.64 \\
 & Replay & 64.17 & 62.18 & 68.00 \\
 & Stage-0 centroid & 73.67 & 66.18 & 65.27 \\
 & Snapshot centroid & 82.50 & 66.00 & 85.09 \\
Return/shadow/tool & Sequential FT & 45.88 & 44.46 & 46.52 \\
 & Replay & 44.31 & 38.31 & 35.00 \\
 & Stage-0 centroid & 55.69 & 55.23 & 48.63 \\
 & Snapshot centroid & 55.88 & 52.00 & 45.46 \\
\bottomrule
\end{tabular}
\end{adjustbox}
\end{wraptable}
\WFclear

\noindent\textbf{Implementation identity and cost.}
The formal rerun uses GRAD's G3 \texttt{ModalityNormGNN} with 512-character
node windows and 160-character edge windows. The pre-run contract tests
pass 18/18, including snapshot save/load identity, replay sampling,
centroid routing, frozen backbone parameters and running statistics, and
the auxiliary-loss and modality-dropout checks. The model has 643,592
parameters (+2.6\%). Stage--0 storage is 2.58\,MB and three snapshots
use 7.74\,MB. For seed 211, batch-one centroid routing takes 7.7--20.1\,ms p50 and
8.3--29.1\,ms p95; all-expert evaluation takes 23.1--58.9\,ms p50 and
23.7--86.1\,ms p95. These measurements are detector inference times;
agent generation and remote service time are excluded.

\noindent\textbf{Training budget and replication.}
All continual-learning arms use 200 updates with development-AUROC
selection and no mining. This differs from the static detection
training recipe. Mean training times across 20 runs are 45 seconds for
sequential fine-tuning, 56 seconds for Replay and stage-0 experts, and
97 seconds for snapshot experts. The GRAD CAA experiment follows the replication timeline in
Appendix~\ref{app:data}. Runner, trainer, permission
specification and stream-manifest hashes accompany the run outputs.

\subsection{Cross-Dataset Gains across Weighting and Resampling Schemes}
\label{app:g3_endpoint_recheck}
The endpoint-analysis source is included in \path{code/project/code/repro/caa_attr_20260923/}; Appendix~\ref{app:code_release} describes the package.

This recheck uses the existing final-stage predictions from the ten
formal G3 Decomp/MCP/MRT runs, with seeds 211--215 and two task orders.
CAA improves accuracy over Replay by 13.08 percentage points under
the original endpoint and by 9.05 points with equal family/base-case
weights. Both gains have positive paired confidence intervals.
The analysis uses stored scores and the original classification threshold.

\noindent\textbf{Endpoint definitions.}
Task-macro sample accuracy weights tasks equally after averaging
sample correctness within each task. Pooled accuracy averages all
test samples. Family-equal accuracy averages correctness within
families, then weights families within tasks and tasks equally.
Table~\ref{tab:g3_endpoint_levels} averages these endpoints over ten
runs on the DMM sample splits. The separate family-isolated stream
is analyzed in Appendix~\ref{app:g3_family_weights}.

\par
\begin{wraptable}{r}{\textwidth}
\centering\small
\caption{G3 cross-dataset accuracy under three weighting schemes, in percent.
The original endpoint is listed first. Bold marks the highest value in
each column, including the task-oracle reference.}
\label{tab:g3_endpoint_levels}
\begin{adjustbox}{max width=\linewidth}
\begin{tabular}{lrrr}
\toprule
Method & Task-macro sample & Pooled sample & Family-equal \\
\midrule
Sequential FT & 56.43 & 56.59 & 62.13 \\
Replay & 64.78 & 64.76 & 57.31 \\
Stage-0 centroid & 68.37 & 68.53 & 56.26 \\
Stage-0 all & 63.89 & 63.88 & 55.80 \\
Stage-0 task oracle & 68.31 & 68.47 & 56.44 \\
\textbf{Snapshot centroid (CAA)} & 77.86 & 78.00 & 66.36 \\
Snapshot all & 71.58 & 71.82 & 64.08 \\
Snapshot task oracle & \textbf{79.86} & \textbf{79.94} & \textbf{68.51} \\
\bottomrule
\end{tabular}
\end{adjustbox}
\end{wraptable}
\WFclear

\noindent\textbf{Paired differences and resampling.}
The original analysis averages the two orders within each seed and
bootstraps the five paired seed differences. Its 13.08-point gain and
95\% interval are reproduced. The two-way sensitivity analysis
independently resamples five seeds and two orders, using their crossed
combinations. Both procedures use 2,000 draws.
Table~\ref{tab:g3_endpoint_intervals} reports all three endpoints; the
snapshot-centroid advantage over Replay remains positive under both
resampling schemes.

\par
\begin{wraptable}{r}{\textwidth}
\centering\small
\caption{Snapshot centroid minus Replay on the cross-dataset stream.
Differences and 95\% confidence intervals are in percentage points.}
\label{tab:g3_endpoint_intervals}
\begin{adjustbox}{max width=\linewidth}
\begin{tabular}{lrrr}
\toprule
Endpoint & Difference & Seed-only CI & Seed $\times$ order CI \\
\midrule
Task-macro sample & \textbf{+13.08} & [9.11, 19.10] & [8.44, 21.15] \\
Pooled sample & +13.24 & [9.18, 19.29] & [8.59, 21.41] \\
Family-equal & \textbf{+9.05} & [6.25, 11.91] & [3.03, 15.55] \\
\bottomrule
\end{tabular}
\end{adjustbox}
\end{wraptable}
\WFclear

Family-cluster sensitivity resamples ten runs and families within
runs, retaining paired sample differences for each selected family.
Across 1,000 draws, the sample-pooled difference has a 95\% interval
of [7.72, 19.94] points. This statistic divides summed differences
by the number of sampled examples; it uses pooled rather than
task-macro weights.

Weighting also changes baseline comparisons. Sequential FT has higher
family-equal accuracy than Replay, 62.13\% versus 57.31\%, despite
the reverse sample-level ranking. Stage-0 centroid minus Replay
changes from +3.59 points under the original endpoint, with a
seed-only 95\% interval of [-1.36, 10.18], to -1.05 points under
family-equal weighting, with an interval of [-4.13, 2.03]. Neither
stage-0 difference excludes zero.

\noindent\textbf{Task and family contributions.}
Table~\ref{tab:g3_endpoint_tasks} shows that Decomp and MRT drive the
aggregate improvement under both weights. MCP changes sign under
family weighting; its interval includes zero. Averaging each MCP
family's difference over runs in which it appears and then weighting
the distinct families equally gives -9.33 points. That diagnostic
differs from the -6.72-point endpoint, which weights families within
each run before averaging runs.

\par
\begin{wraptable}{r}{0.62\textwidth}
\centering\footnotesize\setlength{\tabcolsep}{3pt}
\caption{Per-task snapshot-centroid gains over Replay, in percentage points.
Intervals use the original seed-only paired bootstrap.}
\label{tab:g3_endpoint_tasks}
\begin{adjustbox}{max width=\linewidth}
\begin{tabular}{lcc}
\toprule
Task & \makecell{Sample gain\\{}[95\% CI]} & \makecell{Family-equal gain\\{}[95\% CI]} \\
\midrule
Decomp & \makecell{+18.33\\{}[9.83, 27.83]} & \makecell{+18.38\\{}[9.76, 27.87]} \\
MCP & \makecell{+3.82\\{}[0.00, 7.64]} & \makecell{-6.72\\{}[-17.01, 1.38]} \\
MRT & \makecell{+17.09\\{}[7.64, 26.55]} & \makecell{+15.49\\{}[5.74, 25.24]} \\
\bottomrule
\end{tabular}
\end{adjustbox}
\end{wraptable}

Size sensitivity removes the $\max(1,\lfloor K/20\rfloor)$ largest
of $K$ families per run. The remaining sample-level gains are 18.66,
2.56, and 14.86 points for Decomp, MCP, and MRT. For MRT, this removes
one of five families (20\%). Decomp and MRT retain positive gains;
their largest single-family contributions are 1.7 and 5.8 points.
A contribution is the family sample fraction times its paired accuracy
difference, averaged over runs containing that family.

\par\WrapTableClear
\noindent\textbf{Family definitions and split overlap.}
For Decomp, the numeric case identifier groups harmful and benign
counterparts with the mixture variants of the same base case. There
are 56 test families per run on average, with a mean size of 1.07;
25--35\% of test samples belong to a case also present in training.
For MCP, generated pairs use \texttt{pair\_id}, while runtime examples
use the task template after removing the instance suffix. There are
25.8 test families per run on average, with mean size 2.15; about
60--75\% of test samples share a family with training. MRT uses the
scenario identifier, with five test families of mean size 11; all
five also occur in training.

DMM evaluates task adaptation and retention on test samples drawn from
these task distributions, including recurring attack families. The grouping
units differ: Decomp groups variants of a base case, MCP combines generated
pairs and runtime templates, and MRT groups scenarios. Base-case overlap
therefore differs from sharing only a broad attack strategy. Family-equal
accuracy gives these groups equal weight within each task; the separate
mechanism stream evaluates transfer across held-out families. Cluster
resampling models within-family dependence and does not alter split
membership. Only five MRT scenarios and two orders are available for
the corresponding sensitivity analyses.

\subsection{Family Weighting in the Mechanism Stream}
\label{app:g3_family_weights}
On the ten G3 mechanism-stream runs, the family-equal endpoint
weights label-pure families within tasks and tasks equally, excluding
zero or one mixed-label family per panel-task. Snapshot centroid,
Replay, and stage-0 centroid achieve 47.5\%, 53.5\%, and 50.1\%,
compared with sample-level accuracies of 51.11\%, 39.21\%, and 53.18\%.
Snapshot minus Replay is -6.0 points (run-bootstrap 95\% CI:
-9.7-- -2.2); stage-0 minus Replay is -3.5 points (95\% CI:
-8.5--2.0).

\section{Matched Decision-Time Detection Comparison}
\label{app:w1_comparison}
Source and offline reanalysis commands are listed in Appendix~\ref{app:code_release} and \path{code/CODE_INDEX.md}.

\paragraph{Protocol.}
The obs1 comparison uses the original group-disjoint Decomp fold0 split:
1,397 training, 352 development, and 232 test cases (174 attacks and
58 benign cases in 58 base-case groups). Attacks are truncated through
the proposed harmful action, before its execution; later operations are
excluded. The same 200-character step strings are supplied to all learned
arms. Training uses seeds 211--213, at most 60 epochs, patience 15,
batch size 48, and learning rate $10^{-3}$. Development selection uses
F1 under a target FPR of 0.10. Training ran on RTX 3090 GPUs with the
deterministic entry point. This is a revised comparison on a previously
exposed test split, rather than an independent confirmation.

\paragraph{Single-model and ensemble endpoints.}
Table~\ref{tab:w1_decomp} averages AUROC across the three separately
trained models; its standard deviation uses denominator three.
The corresponding mean GRAD improvements are 5.30 points over Temporal
and 10.64 points over Text-only. Table~\ref{tab:w1_paired} instead reports
AUROC differences after averaging the three models' scores for each
case. Its 95\% intervals resample 58 base-case groups (5,000 draws).
These ensemble intervals do not describe the mean single-model differences.

\par
\begin{wraptable}{r}{\textwidth}
\centering\small
\caption{Score-ensemble AUROC differences (percentage points), with
95\% base-case cluster bootstrap intervals.}
\label{tab:w1_paired}
\begin{tabular}{lrr}
\toprule
Comparison & Difference & 95\% CI \\
\midrule
GRAD $-$ Temporal & +4.85 & [+1.32, +9.08] \\
GRAD $-$ Text-only & +6.40 & [+2.48, +10.99] \\
Temporal $-$ Text-only & +1.56 & [$-3.13$, +6.29] \\
\bottomrule
\end{tabular}
\end{wraptable}
\WFclear

For individual seeds 211/212/213, GRAD--Temporal differences are
3.06/7.07/5.77 points with intervals [$-1.95$,8.21], [2.75,11.84],
and [2.13,9.74]. GRAD--Text-only differences are 7.67/10.99/13.26
points with intervals [2.47,13.41], [5.28,17.20], and [7.50,19.38].
Temporal--Text-only intervals exclude zero in one of three seeds;
the score-ensemble interval includes zero.

\paragraph{Paired mean-AUROC analysis and frozen operating points.}
The W4 reanalysis resamples the same 58 base-case groups jointly across
methods and seeds, recomputing each seed's AUROC and then averaging
paired differences within each draw. The 95\% intervals for mean
GRAD--Temporal and GRAD--Text-only differences are [2.0,9.1] and
[7.2,14.4] percentage points. This analysis uses the existing exposed
Decomp test set and the same frozen model scores.
For the separate score-ensemble endpoint, the updated intervals are
[1.3,9.0] and [2.4,11.1] points, respectively; Table~\ref{tab:w1_paired}
retains the original W1 bootstrap realization.

\par
\begin{wraptable}{r}{\textwidth}
\centering\small
\caption{W4 test TPR/FPR (\%) under development-frozen FPR targets.
Values are means over seeds 211--213; targets apply to development data.}
\label{tab:w4_points}
\begin{tabular}{lcc}
\toprule
Method & Dev target 5\%: TPR/FPR & Dev target 10\%: TPR/FPR \\
\midrule
GRAD & 85.6/7.5 & 96.0/16.7 \\
Temporal & 67.8/5.2 & 77.6/13.2 \\
Text-only & 54.4/9.2 & 78.2/20.1 \\
\bottomrule
\end{tabular}
\end{wraptable}
\WFclear

Thresholds maximize development recall subject to the FPR target;
unique candidates are traversed in descending order and decisions use
score $\geq$ threshold. This differs from the original W1 threshold
selection in Table~\ref{tab:w1_decomp}. Development and test contain
88 and 58 benign cases per seed, giving FPR steps of 1.14 and 1.72
percentage points. Test FPR exceeds both nominal development targets
on average for all three methods. At either target, GRAD has higher
TPR and lower FPR than Text-only, whereas Temporal has lower FPR and
lower TPR than GRAD. These are transferred operating points, not
comparisons at equal test FPR; 1\% operating points are not reported.

\paragraph{Monitor and cost.}
The sequential monitor reuses cumulative binary judgments for all
232 cases from 2,267 cached DeepSeek calls (historical cost CNY 0.4671;
no additional calls). Applying the decision-time prefix gives
TP/FP/TN/FN = 174/22/36/0. An ordinal score derived from first-trigger
position has AUROC 91.24\%; it is distinct from a native continuous
risk score. Length alone has AUROC 77.70\%, so this comparison does
not establish invariance to trace length or isolate the causal effect
of graph edges. It evaluates detectors, not full reproductions of
external defense pipelines.
On an M2 CPU with batch size one, GRAD/Text-only/Temporal inference
p50 is 6.8/2.9/2.6\,ms and p95 is 45.5/4.1/4.1\,ms. Parameter counts
are 643,592/623,298/673,090 and checkpoints occupy 2.60/2.54/2.73\,MB.
These CPU measurements are separate from the server gate/hosted-monitor
measurements in the closed-loop experiment.

\paragraph{Original observation protocols.}
Tables~\ref{tab:detection} and~\ref{tab:binary_checking} in this supplement
report the original Decomp full-trace and MCP/MRT comparisons.
Their observation and threshold protocols are specified in
Appendix~\ref{app:detection_versions}.

\clearpage
\section{Detailed AI Use Statement}
\label{app:ai_use}

We disclose five uses of generative AI in preparing this work:
writing assistance, literature retrieval and discovery, research and
experimental execution, manuscript drafting, and synthetic dataset generation. These categories correspond
to the AI-assistance disclosure submitted with the paper.

\subsection{Writing Assistance and Manuscript Drafting}
Generative AI assisted with language editing, polishing, and LaTeX
preparation. This included revising wording for clarity, reducing
repetition, organizing paragraphs, and adjusting manuscript presentation.
AI assistance also supported figure preparation, vector reconstruction,
and supplementary-artifact organization; these presentation changes did
not introduce new experimental measurements.
AI tools also helped draft sections from the research materials and
experimental summaries supplied by the authors. The authors revised
these drafts and selected the text included in the manuscript.
Writing assistance and section drafting are disclosed separately because
the assistance covered both revision of existing prose and preparation
of new draft passages.

\subsection{Literature Retrieval and Discovery}
AI tools assisted in identifying relevant literature and organizing
related-work discussions. This use supported discovery of candidate
references and connections to the topics discussed in the paper.
Bibliographic entries were initially obtained from Google Scholar exports;
identified metadata errors were corrected using official proceedings.
Source attribution is retained in the manuscript. The authors are
responsible for the selection of references and the accuracy of the
claims attributed to them.

\subsection{Research and Experimental Execution}
Generative AI tools also assisted with research and experimental
execution. This assistance extended beyond manuscript preparation and
is disclosed as a separate use. The authors take responsibility for
the research procedures, experimental artifacts, and interpretation of
results. The methods, protocols, and archived outputs described in this
supplement provide the basis for evaluating the reported findings.

\subsection{Synthetic Dataset Generation}
Generative AI was used in preparing synthetic datasets for the study.
The data descriptions separate benchmark-derived material from
synthetic examples, with construction and evaluation procedures
reported in Appendix~\ref{app:data} and the corresponding experimental
protocols. AI-generated examples are synthetic material; their generation
alone does not establish that an attack succeeds during agent execution.
Offline detection and closed-loop outcomes are evaluated and reported
under their respective protocols. Language models additionally appear
as agents and monitors in the evaluated systems; those experimental
roles are described alongside the relevant comparisons.

\subsection{Author Responsibility}
The authors take responsibility for the final text, reference selection,
datasets, and interpretation of the results. Reported numerical findings
are based on recorded experimental outputs and the aggregation procedures
described in the paper. AI-assisted prose and summaries do not replace
those experimental records. The reproducibility documentation provides
the protocol and result details needed to assess the reported findings.

\clearpage
\section{Supplementary Code and Artifact Guide}
\label{app:code_release}

The companion code-and-evidence archive contains a \path{code/}
folder. All paths below are relative to that archive root.
\path{code/README.md} gives installation and offline verification commands;
\path{code/CODE_INDEX.md} maps the experiments to their source files.
The internal project layout is preserved under \path{code/project/}.
The launcher \path{code/run.py} selects that project root automatically.

\subsection{Model and Experiment Entry Points}
The SSUM trainers and encoder ablations are in
\path{code/project/code/repro/final_five_tables/}, including
\path{train_grad_repair.py} and \path{train_gnn_ablate2.py}.
The matched-observation data builder, GRAD/text/temporal wrappers,
and aggregation are in
\path{code/project/code/repro/w1_fair_comparison/}.
W4 paired statistics and development-frozen operating points are
implemented in
\path{code/project/code/repro/w1_final_confirmation/final_stats.py}
with the \path{--target test} argument.

CAA snapshot training and routing are implemented in
\path{code/project/code/repro/real_cl_g3.py}; endpoint and family-weight
analyses are in \path{code/project/code/repro/caa_attr_20260923/}.
The gate, transfer training, online runner, budget client, aggregation,
and contract tests are in \path{code/project/code/repro/final_table4/}.
AgentDojo diagnostics are in \path{code/project/code/repro/final_table2/}.
Historical baseline adaptations and their dependencies are identified
separately in \path{code/CODE_INDEX.md}.

\subsection{Artifacts, Portability, and Verification}
Selected recorded outputs accompany the source under
\path{code/project/LOGS/}. These include the W1 per-example predictions,
W4 statistics, the earlier and T3 closed-loop panels, and CAA summaries.
The package includes W1 and gate checkpoints with anonymized string
metadata. Their model tensors were verified exactly equal to the
archived originals; reserialization changes whole-file hashes.
\path{code/source_manifest.json} records both original and packaged
SHA-256 values, while \path{code/PACKAGING_CHANGES.md} records portability
and anonymization changes. Historical hashes in experiment records refer
to the original executed artifacts.

\path{code/verify_bundle.py} checks packaged-file integrity, Python syntax,
and the supplied artifact paths. The README includes commands for
recomputing the 240-run T3 aggregate and the W4 test statistics without
API calls. Full training and optional historical integrations additionally
require the upstream data and runtimes listed in
\path{code/DATA_AND_DEPENDENCIES.md}. The release includes selected
reanalysis inputs, rather than the entire experimental workspace.

\clearpage
\section{Frozen New-Case Validation and Relation Attribution}
\label{app:new_validation}
The following checks use frozen W1 models and distinguish new generated
cases from attribution on the original exposed split. The source reports
and numerical outputs are included in
\path{code/project/LOGS/core_review_resolution/}.

\paragraph{New generated cases.}
After freezing nine W1 checkpoints, preprocessing, decision-time windows,
and development thresholds, 60 new base tasks with matched benign
counterparts were generated once. Rule-based checks and an independent
GLM validity assessment excluded 13 tasks before model scoring, leaving
47 base-case clusters and 188 examples. These are LLM-generated cases
scored offline, not executed real-world trajectories.
GRAD, Temporal, and Text-only attain seed-mean AUROCs of 0.615, 0.675,
and 0.583. The paired GRAD--Temporal difference is $-0.060$
(95\% CI: [$-0.115,-0.0001$]); the GRAD--Text-only difference is
$+0.032$ ([$-0.021,+0.089$]). Thus, the original advantage over
Temporal does not reproduce. At the frozen 5\% development FPR target,
GRAD's test TPR/FPR is 0.435/0.340. Thresholds were not recalibrated
on these cases.

\paragraph{Candidate-relation attribution.}
On the original W1 split, candidate relations (E), no cross-node edges
(Z), and type/edge-count/degree-matched random rewiring (R) yield
seed-mean AUROCs of 0.970, 0.898, and 0.959. E--Z is $+0.072$
(95\% CI: [$+0.039,+0.110$], Holm $p=0.0004$); E--R is $+0.011$
([$-0.003,+0.026$], Holm $p=0.126$). Candidate relations provide
an increment over removing cross-node edges, but their specific wiring
has no established advantage over the matched rewiring control.
These results do not establish recovery of causal execution dependencies.
The E wrapper reproduces the frozen weights and per-example predictions
bitwise across all three seeds. CAA was not retrained or evaluated in
these checks.

\end{document}